\documentclass[11pt,tightenlines,eqsecnum,floats,aps,amsmath,amssymb,nofootinbib,prd,shownopacs,floatfix]{revtex4-2}
\usepackage{comment}

\usepackage{graphicx}

\usepackage{epstopdf}
\usepackage{latexsym}
\usepackage{amssymb}
\usepackage{amsmath}
\usepackage{color}
\usepackage{mathrsfs}
\usepackage{xparse}
\usepackage{float}
\usepackage{mathtools}
\usepackage{multirow}

\usepackage[table,xcdraw]{xcolor}

\usepackage[normalem]{ulem}
\useunder{\uline}{\ul}{}

\usepackage[center]{subfigure}

\begin{document}
\renewcommand\arraystretch{2}
 \newcommand{\bq}{\begin{equation}}
 \newcommand{\eq}{\end{equation}}
 \newcommand{\bqn}{\begin{eqnarray}}
 \newcommand{\eqn}{\end{eqnarray}}
 \newcommand{\nb}{\nonumber}

 \newcommand{\cb}{\color{blue}}

    \newcommand{\cc}{\color{cyan}}
     \newcommand{\lb}{\label}
        \newcommand{\cm}{\color{magenta}}
\newcommand{\rc}{\rho^{\scriptscriptstyle{\mathrm{I}}}_c}
\newcommand{\rd}{\rho^{\scriptscriptstyle{\mathrm{II}}}_c}
\NewDocumentCommand{\evalat}{sO{\big}mm}{%
  \IfBooleanTF{#1}
   {\mleft. #3 \mright|_{#4}}
   {#3#2|_{#4}}
}

\newcommand{\PRL}{Phys. Rev. Lett.}
\newcommand{\PL}{Phys. Lett.}
\newcommand{\PR}{Phys. Rev.}
\newcommand{\CQG}{Class. Quantum Grav.}
\newcommand{\parallelsum}{\mathbin{\!/\mkern-5mu/\!}}

\title{Ekpyrosis in Quantum Gravitational Anisotropic Bouncing Models}
\author{Rachel Brown $^{1}$}
\email{rbro169@lsu.edu}
\author{A. Meenakshi McNamara $^{2,3}$}
\email{amcnamara@perimeterinstitute.ca}
\author{Sahil Saini $^{4}$}
\email{sahilsaiini@gjust.org}
\author{Parampreet Singh $^{1,5}$}
\email{psingh@lsu.edu}
\affiliation{$^{1}$ Department of Physics and Astronomy, Louisiana State University, Baton Rouge, LA 70803, USA\\$^{2}$ Purdue University, Mathematical Sciences Bldg, 150 N University St, West Lafayette, IN 47907\\
$^{3}$ Perimeter Institute for Theoretical Physics,
51 Caroline Street, Waterloo, Ontario, Canada\\
$^{4}$ Department of Physics, Guru Jambheshwar University of Science $\&$ Technology, Hisar, Haryana 125001, India\\
$^{5}$ Center for Computation and Technology, Louisiana State University, Baton Rouge, LA 70803, USA}

\begin{abstract}

We explore the isotropization of \textcolor{black}{ a model anisotropic} universe 
in the bouncing models using the ekpyrotic \textcolor{black}{potential without assuming initial conditions corresponding to an ekpyrotic phase.}  \textcolor{black}{In particular, we explore the way use of ekpyrotic potentials may dynamically help isotropization for the considered initial conditions corresponding to the macroscopic classical contracting universe with potentially large anisotropies.} 
As an example of a concrete non-singular bouncing mechanism, we consider the effective description of loop quantum cosmology for Bianchi-I and Bianchi-IX spacetimes for ekpyrotic and ekpyrotic-like potentials. \textcolor{black}{Considering two different values of potential parameters and initial conditions corresponding to a classical macroscopic universe,}  we show that for both of these spacetimes the cosmological singularity is resolved via multiple short-duration non-singular bounces \textcolor{black}{caused by quantum gravitational effects.} 
\textcolor{black}{We perform a large number of numerical simulations for a wide range of initial conditions which do not favor ekpyrosis initially. 
Even with such unfavorable initial conditions we show 
 that the relative strength of the anisotropies at the end of the bounce regime is noticeably reduced in more than $90\%$ of the simulations.
 This provides a strong evidence for the isotropization ability of the ekpyrotic potentials.} 
 \textcolor{black}{We find that 
isotropization can occur} over cycles of rapid non-singular bounces in the Planck regime via enhancement of the contribution of the (isotropic) energy density relative to the anisotropies at the bounces. 
Achieving isotropization is found to be easier in Bianchi-I spacetimes when compared to Bianchi-IX spacetimes. \textcolor{black}{Our results demonstrate that, even with initial conditions which are not most favorable for existence of ekpyrosis, an effective isotropization can occur in non-singular anisotropic models with ekpyrotic and ekpyrotic-like potentials. }
\end{abstract}

\maketitle

\section{Introduction}

In recent years, various investigations have been carried out on bouncing cosmologies to determine whether they can serve as potentially viable alternatives to inflation. A typical scenario in these models consists of a \textcolor{black}{contracting macroscopic universe from the classical regime that undergoes a bounce, from which an expanding universe emerges.} 
The first challenge in constructing such bouncing models is the cosmic singularity which is inevitable in general relativity (GR) unless the null energy condition is violated which often results in pathologies. For a viable bouncing scenario, it is required that the singularity resolution be a generic phenomena and not a result of fine-tuning of initial conditions or the choice of some exotic matter. If the fundamental challenge of obtaining singularity resolution is addressed, bouncing models face the next challenge -- the growth of anisotropies in the contracting branch. As the mean scale factor decreases, the anisotropic shear grows at a faster rate than the energy density of matter source with an equation of state $w < 1$. As a result,  the approach to big bang/big crunch singularities is dictated by anisotropies for such a matter source resulting in a mixmaster dynamics and BKL instability \cite{Belinsky:1970ew,belinskii1970oscillatory,misner1969mixmaster} which can potentially also affect the viability of a non-singular bounce. Even in the simplest anisotropic models such as Bianchi-I spacetime, anisotropies dominate the approach to singularity \cite{jacobs1968spatially}. Thus, even if a bounce occurs, it is expected to be highly anisotropic and the resulting post-bounce universe may also retain this character. The pertinent question is whether one can construct robust non-singular bouncing models in anisotropic spacetimes where anisotropies are tamed in the contracting branch, allowing the universe to isotropize.

To answer this question, one needs two main ingredients: a mechanism to resolve the singularities in the anisotropic setting and a mechanism to diminish the anisotropies. Let us first consider the resolution of singularities. It has long been expected that to robustly address the problem of cosmological singularities, one would need some quantum gravitational input. One of the arenas where this has been rigorously demonstrated is loop quantum cosmology (LQC) \cite{Ashtekar:2011ni} where cosmological singularities have been demonstrated to be generically resolved without imposing any fine-tuned initial conditions or any violation of energy conditions \cite{Singh:2014fsy, Li:2023dwy}. In LQC, the big bang singularity no longer exists and an expanding universe emerges from a contracting universe which experiences a bounce caused by non-perturbative quantum geometry effects which dominate the dynamics when the energy density reaches Planckian values during contraction \cite{Ashtekar:2006rx, Ashtekar:2006wn}. \textcolor{black}{To understand singularity resolution and the underlying physics at the Planck scale in LQC, one considers initial states in a macroscopic classical universe and evolves them towards classical big bang using quantum Hamiltonian constraint. As one approaches, Planck regime there are departures from the classical Einstein's equations due to quantum gravitational effects which result in a bounce when the spacetime curvature becomes Planckian.} This result has been found to be robust for various spacetimes \cite{Ashtekar:2007em}, including in the presence of spatial curvature \cite{Ashtekar:2006es, Szulc:2006ep, Vandersloot:2006ws}, cosmological constant \cite{Kaminski:2009pc, Pawlowski:2011zf}, anisotropies \cite{Martin-Benito:2009xaf,Diener:2017lde} and Fock quantized inhomogeneities \cite{Brizuela:2009nk}. \textcolor{black}{The singularity resolution in LQC occurs because of the underlying quantum geometry effects which lead to maximum allowed values of energy density and anisotropic shear \cite{Corichi:2009pp}.} The results of singularity resolution have also been extended to black hole spacetimes (see \cite{Ashtekar:2023cod} for a review). 
These results indicate that the problem of singularities is robustly addressed in LQC, and for this reason we will consider this as a concrete example for a non-singular bouncing mechanism in this manuscript.

The advantage of choosing LQC is that a fair amount of work has been done to understand the phenomenology of the anisotropic models in LQC. While the underlying geometry in LQC is discrete and is described by difference equations, for phenomenological explorations it is common to use the continuum effective description of LQC, which is based on differential equations and is shown to faithfully incorporate the underlying quantum geometry corrections in various isotropic and anisotropic spacetimes including the Bianchi-I model \cite{Diener:2013uka,Diener:2014mia,Diener:2014hba,Diener:2017lde}. \textcolor{black}{The validity of effective description is found using extensive numerical simulations where as described above one starts from a macroscopic classical universe.} Using effective spacetime description, where one can express quantum gravitational modifications in terms of modifications to the classical Hamiltonian \cite{Singh:2006sg, Ashtekar:2006wn}, it has been shown that irrespective of the choice of the matter content all strong curvature singularities are generically resolved, and isotropic and anisotropic cosmological spacetimes are geodesically complete \cite{Singh:2009mz, Singh:2010qa, Singh:2011gp, Saini:2016vgo, Saini:2017ipg, Saini:2017ggt, Saini:2018tto}. 
The energy density and anisotropic shear are bounded in Bianchi-I, II and IX spacetimes \cite{Ashtekar:2009vc,Corichi:2009pp,Gupt:2011jh, Singh:2013ava,Wilson-Ewing:2010lkm}.   The Bianchi-I and Bianchi-IX models in LQC have an especially rich phenomenology of anisotropic Kasner transitions at the bounce \cite{Gupt:2012vi,Wilson-Ewing:2017vju,Wilson-Ewing:2018lyx,Blackmore:2023wiv}, \textcolor{black}{whose physical implications have been studied (see for eg. \cite{Gupt:2013swa,Linsefors:2014tna, Motaharfar:2023hil, Motaharfar:2024afn}). Note that in our manuscript, we focus on anisotropies in the context of bouncing models in LQC but there have been various other studies in the context of pre-big bang scenarios, see for eg. \cite{Gasperini:2002bn,Gasperini:2007vw}.}


As far as the issue of suppression of anisotropies is concerned, perhaps the most straightforward way is to consider a source of matter-energy which an equation of state $w > 1$. In this case the energy density of matter can grow faster than anisotropic shear and it is possible that the universe can isotropize as the scale factor decreases. \textcolor{black}{Such an equation of state can occur in ekpyrotic scenarios \cite{Khoury:2001wf, Steinhardt:2002ih} where it is possible to have $w \gg 1$, a condition necessary for  ekpyrosis. In this manuscript, we would label such an equation of state as ultra-stiff.} \textcolor{black}{These models were originally motivated by bulk-brane dynamics where inter-brane dynamics is captured by the evolution of moduli field in a negative exponential potential. The dynamics of the ekpyrotic field results in different phases of evolution of the universe. In the ekpyrotic potential (see eq. (3.1)) when the field is in the large positive regime and moving towards smaller $\phi$ values the universe is classical and macroscopic and in a contracting phase. The field then enters a region of negative well in the potential where conditions favorable for ekpyrosis can arise. As the field approaches large negative values, branes come close to each other and eventually collide causing a big crunch. 
After the collision of branes, the field moves from large negative values towards positive values and the universe enters an  expanding phase from a small size to macroscopic values. Subsequently, the field reaches positive values, and the universe becomes macroscopic, Hubble friction stops the growth of the field causing it to turn around and the cycle repeats. Note that 
as the branes move towards a collision (or come out of a collision), the moduli field can have a pressure greater than the energy density and the ekpyrotic field behaves as an isotropic fluid having an equation of state larger than unity. Our goal in this manuscript is to study the evolution of macroscopic contracting universe starting with large anisotropies in ekpyrotic and ekpyrotic-like potentials without assuming any conditions which may initially favor or guarantee ekpyrosis. In particular, we consider initial conditions corresponding to the classical macroscopic universe with the field taking positive values initially and moving towards negative values. In this way, the considered initial conditions correspond to starting from the contracting phase of the macroscopic universe as described above.  As we show in our analysis, even for not so favorable initial conditions, dynamical evolution results in ekpyrosis at a later stage in evolution causing isotropization.} 

Recall that in the ekpyrotic phase, the energy density grows faster than anisotropies as the universe contracts towards the singularity potentially leading to its isotropization and potentially avoiding the BKL instability \cite{Erickson:2003zm, Garfinkle:2008ei}. Recently, ekpyrotic mechanisms have been used in models with ultra-slow contraction to explain homogeneity, flatness and isotropy in the contracting branch \cite{Ijjas:2019pyf,Ijjas:2024oqn}.  Models which are ekpyrotic-like, with a different negative potential, can also potentially allow an isotropization \cite{Cai:2012va, Cai:2013vm}. If the ekpyrotic field is combined with a bounce, such as in LQC, one can expect the bounce to isotropize, at least when one starts with anisotropies which are small. 
While the intuitive reason for the ekpyrotic field to diminish the growth of anisotropies in the contracting branch is easy to understand, whether or not sufficient and effective ekpyrosis occurs depends on the details of the background dynamics and is potentially sensitive to the initial conditions, including the magnitude of initial anisotropic shear compared to matter energy density. Therefore an important question with any ekpyrotic mechanism is its duration and strength and whether it can result in isotropization for arbitrary initial conditions, especially if the anisotropies are large. \textcolor{black}{An example of this occurs in bouncing models in LQC where ekpyrotic potentials have been considered in both isotropic and anisotropic models showing that a non-singular evolution is possible. 
Answering the above question on the duration and strength of ekpyrosis is important} to construct viable models of the early universe where curvature perturbations start in the contracting branch before the bounce, such as in the matter-bounce alternative to inflation \cite{Finelli:2001sr,Brandenberger:2009yt,Cai:2011tc,Li:2020pww}. 
\textcolor{black}{
Since we consider initial conditions corresponding to the macroscopic classical contracting universe,  the numerical simulations performed in this manuscript are only for the initial conditions when the field is in the positive regime and moving towards smaller $\phi$ values. This does not guarantee that ekpyrosis occurs for considered initial conditions and whether or not the universe undergoes ekpyrosis is determined by the dynamical evolution. An advantage of considering these initial conditions is two fold. First, these initial conditions are imposed in a classical macroscopic universe starting far from the quantum gravitational regime which allows to study and compare evolution of different quantum anisotropic universes with similar initial conditions. And second, by considering such initial conditions we are exploring isotropization in a more difficult setting where one does not  assume favorable initial conditions for ekpyrosis.}\footnote{\textcolor{black}{In contrast, one can always assume initial conditions which favor ekpyrosis, such as starting in the regime when field takes negative values. However, our goal in this paper is to consider initial conditions which do not start with ekpyrosis and let dynamics determine whether isotropization occurs.}}

The goal of this manuscript is to understand whether \textcolor{black}{starting from a macroscopic classical universe with an ekpyrotic potential and the scalar field rolling down from the positive values, there can be an effective isotropization of the non-singular bounce. } We consider 
Bianchi-I and Bianchi-IX models in the effective spacetime description of LQC as our setting and contrast the dynamics when the matter-energy is sourced by a  massless scalar field only versus the ekpyrotic field both in the ekpyrotic potential \cite{Steinhardt:2002ih} as well as an ekpyrotic-like potential \cite{Cai:2012va}. Specifically, we study whether the ekpyrotic field can dominate the bounce regime and lead to a reduction in the strength of the anisotropic shear relative to the energy density of the universe (which is assumed to be isotropic). \textcolor{black}{It is important to note that we neither make any assumptions about the existence of ekpyrotic phase nor we consider specific initial conditions which favor existence of such a phase. Rather, the initial conditions we consider correspond to a contracting macroscopic universe from classical scales towards the big crunch, specifically when the field starts from positive values and rolls towards negative values in the ekpyrotic potential. The goal of this work is not to construct a viable ekpyrotic model but to understand the way ekpyrotic potential might help in isotropization.} Earlier studies with ekpyrotic potential in LQC confirm that non-singular cyclic evolution is possible
\cite{Bojowald:2004kt,Singh:2006im}, but it is only in the presence of anisotropies \cite{Cailleteau:2009fv} that the moduli field can turn around which is necessary for a viable ekpyrotic scenario as proposed originally \cite{Khoury:2001wf}. Similarly, attempts have been made to explore the effect of anisotropic evolution on the cosmological perturbations in LQC, however these studies also restrict anisotropies assuming a point-like bounce and a fixed ultra-stiff equation of state \cite{Agullo:2020wur,Agullo:2022klq}. But it should be noted that 
in general, the anisotropic approach to singularity (and hence the bounce) would be cigar-like with two scale factors contracting and one expanding before the bounce. 
To obtain a point-like bounce (all scale factors contracting pre-bounce) under generic anisotropic conditions it is necessary to have a viable ekpyrotic scenario. \textcolor{black}{Thus, it is important to understand whether ekpyrosis can isotropize an anisotropic bounce. 
In this manuscript, we go beyond the assumption of small anisotropic perturbations and explore whether an isotropic universe can be obtained after bounce(s) starting from a contracting macroscopic anisotropic universe in the classical regime in the presence of an ekpyrotic potential without making any restrictions on the equation of state.}

The effective equations of motion for the loop quantized  Bianchi-I and Bianchi-IX spacetimes are complicated and efforts are still underway to develop an intuitive understanding of the interplay between the energy density and the anisotropic shear in these models. Recently, a parabolic relation between the anisotropic shear and the energy density has been found to hold at the bounce for certain matter fields in the Bianchi-I model using extensive numerical simulations \cite{McNamara:2022dmf}. This is very surprising in light of the complicated nature of the effective equations and the fact that no such simple relation is found to hold before or after the bounce. However, the current state of the art is far from the level of analytical control needed to go beyond making a few qualitative remarks, and the progress is more often based on numerical simulations. Consequently, our analysis in this manuscript is numerical, yet wide enough to gain general insights on possible outcomes from a matter-ekpyrotic bounce. We consider the effective Bianchi-I and Bianchi-IX spacetimes for ekpyrotic and ekpyrotic-like potentials. \textcolor{black}{In particular, we consider two different values of the strengths of ekpyrotic and ekpyrotic-like potentials and perform $500$ numerical simulations each with varying initial conditions chosen randomly in regime when the universe is initially classical, macroscopic and contracting towards a bounce. Since we are working in the Hamiltonian framework, the phase space variables are three triad components $p_i$, three connection components $c_i$, and the matter phase space variables: the scalar field $\phi$ and its conjugate $p_\phi$. Since we start the numerical simulations in the classical regime we take the initial values of the scalar field and its momentum such that the energy density $\rho$ is in the range $\rho \in [0,10^{-4}]$ in Planck units. Further, the value of $\phi$ is taken to be in the range $\phi \in [0,0.4]$. Given the initial values of $\phi$ and $\rho$, the initial value of $p_\phi$ (or $\dot \phi$) can be determined. Note that the imposition of Hamiltonian constraint reduces one degree of freedom. As a result, we provide initial conditions of three triads and two connection components only in the gravitational sector. The remaining connection component is determined from the Hamiltonian constraint. To measure the extent of the isotropization achieved due to ekpyrotic potentials, we use the massless scalar field as the reference case against which the extent of isotropization is measured. Specifically, each simulation with the ekpyrotic or ekpyrotic-like potentials is accompanied by a corresponding simulation with only the massless scalar field with same initial conditions for three triads, two connection components, energy density and the value of the massless scalar field in the range discussed above. The initial conditions for massless case are obtained by putting potential term to zero while keeping the initial conditions for three triads, two connection components, $\phi$ and $\rho$ to be same. Since the Hamiltonians of massless and non-zero potential are slightly different in the range of energy densities we perform our simulations, the initial value of connection component determined using Hamiltonian constraint for massless and non-zero potential cases is slightly different at the classical scales.} Comparing the relative strength of the anisotropies at the bounce (relative to the matter energy density) we state that isotropization has been achieved if the relative strength of the anisotropies has decreased in comparison with the simulation with massless scalar field with matching initial conditions. 

\textcolor{black}{
Due to quantum geometry effects there are theoretical maximum values of energy density and anisotropic shear, given respectively by $\rho_{\mathrm{max}} \approx 0.41 \rho_{\mathrm{Pl}}$ and $\sigma^2_{\mathrm{max}} \approx 11.57 l_{\mathrm{Pl}}^{-2}$. As a result of these bounds,  we find that even though the initial values of $\phi$ and $\dot \phi$ are not large enough to yield a long range of ekpyrosis, 
isotropization is achieved in an overwhelming majority of the simulations, in both the Bianchi-I and Bianchi-IX models when the ekpyrotic field is present. Even though  the equation of state of the ekpyrotic fields becomes ultra-stiff only for a small duration near the bounce, isotropization can occur.} \textcolor{black}{We find that in ekpyrotic potentials, \textcolor{black}{ with the value of the field taken as positive initially}, a \textcolor{black}{macroscopic} contracting \textcolor{black}{classical} universe generically undergoes multiple non-singular bounces in rapid succession before the emergence of a macroscopic expanding universe, causing multiple short spurts of ekpyrosis.} These short spurts of ekpyrosis coupled with multiple bounces are found to be sufficient to produce isotropization. Another important insight we obtain is that the effect of the ekpyrotic potential on the anisotropic shear is actually indirect. This information is distilled by analyzing the effect of the ekpyrotic field on the energy density and the anisotropic shear, separately, in each simulation. We find that the ekpyrotic potential leads to a direct rise in the energy density at the bounce, indicated by a rise in the energy density at the bounce compared to the massless scalar field case in an overwhelming majority of the simulations. \textcolor{black}{Since the spacetime curvature at the bounce gets contribution both from energy density and anisotropic shear, the relative strength of anisotropies in general decreases in the bounce regime.}
It turns out that the isotropization is achieved in a slightly higher fraction of cases in the Bianchi-I model as compared to the Bianchi-IX model. Further, we find that increasing the strength of the ekpyrotic potentials does result in a higher fraction of cases where isotropization is achieved. These findings are summarized in a table at the end of the manuscript \textcolor{black}{for 4000 numerical simulations} which provides a more granular look at these results. 

The manuscript is organized as follows. In section II, we briefly describe the effective dynamics of Bianchi-I and Bianchi-IX spacetimes in LQC and revisit the general aspects of the ekpyrotic and ekpyrotic-like potentials in these effective spacetimes. Here we demonstrate singularity resolution and some generic features. In section III, we consider the details of ekpyrosis and isotropization in Bianchi-I effective spacetime. Both the ekpyrotic and ekpyrotic-like potentials are considered in the Bianchi-I effective spacetime with two different levels of the strength of the potential. The methodology as well as the initial conditions are also described in this section and various figures summarizing the results of the simulations are included. In section IV, a similar numerical analysis for the Bianchi-IX effective spacetime is presented.  Finally, we summarize our findings in section V.

\section{Effective dynamics of Bianchi-I and Bianchi-IX models in LQC}
In this section, we summarize the basic equations in the classical and effective description of the Bianchi-I and Bianchi-IX models.  
 Starting with the classical dynamics in the canonical framework using Ashtekar-Barbero variables, we briefly describe the equations of motion for effective Bianchi-I and Bianchi-IX spacetimes in LQC. 

\subsection{Classical dynamics of Bianchi-I model}

We consider here the diagonal Bianchi-I model, which under a homogeneous slicing has the topology $\Sigma \times \mathbb{R}$, where the hypersurfaces $\Sigma$ are flat with topology $\mathbb{R}^3$. The metric with lapse $N=1$ is given by
\begin{equation}
    \mathrm{d} s^2=-\mathrm{d}t^2+a_1^2(t) \mathrm{d}x^2 + a_2^2(t) \mathrm{d}y^2 + a_3^2(t) \mathrm{d}z^2,
\end{equation}
where $a_i(t)$ are the directional scale factors. The metric (2.1) reduces to the Friedmann-Lemaitre-Robertson-Walker (FLRW) metric in the isotropic limit, which describes a homogeneous, spatially flat, and isotropic universe. In the classical theory, as well as in the loop quantization of this model, as considered in \cite{Ashtekar:2009vc} we can consider spatial topologies which are both compact ($\mathbb{T}^3$) and non-compact ($\mathbb{R}^3$) \cite{Corichi:2009pp}. If the choice of spatial topology is non-compact, it is necessary to include a fiducial cell to describe the symplectic structure on the phase space. In this manuscript, we consider the spatial topology to be non-compact, and we take the coordinate volume of the fiducial cell to be unity. The  Ashtekar-Barbero connection $A^i_a$ and the triads $E^a_i$ when symmetry reduced, have one independent component per spatial direction, which we denote as connections $c^i$ and triads $p_j$ respectively. They satisfy the following Poisson brackets
\begin{equation}
    \lbrace c^i,p_j \rbrace = 8\pi G \gamma \delta^i_j,
\end{equation}
where $\gamma\approx0.2375$ is the Barbero-Immirizi parameter whose value is set from black hole thermodynamics in LQG. The triads are given in terms of the scale factors as
\begin{equation}
    p_1=|a_2 a_3|, \quad p_2=|a_1 a_3|, \quad p_3=|a_1 a_2|.
\end{equation}
The modulus sign is a consequence of the triad orientation. For our analysis, we take this sign to be positive without limiting generality. 
We first briefly discuss the classical dynamics of the Bianchi-I model. The Hamiltonian constraint is given by:
\begin{equation}
    \mathcal{H_\mathrm{cl}^{\mathrm{I}}}=-\frac{1}{8\pi G \gamma^2 v} \left(c_1 p_1 c_2 p_2 + \textit{cyclic} \right) + \mathcal{H_\mathrm{m}} \approx 0 .
\end{equation}
 Here $\mathcal{H_\mathrm{m}}$ is the matter component, 
and $v=\sqrt{p_1 p_2 p_3}=a_1 a_2 a_3$ denotes the physical volume of a unit comoving cell. We obtain the dynamical equations for the triads and connections by applying Hamilton's equations to the classical Hamiltonian constraint as follows: 
\begin{equation}
\dot{p_i} = \lbrace p_i,H_{\mathrm{cl}} \rbrace = -8 \pi G \gamma \frac{\partial\mathcal{H}_{\mathrm{cl}}}{\partial c_i}  \quad \text{and} \quad \dot{c_i} = \lbrace c_i, H_{\mathrm{cl}} \rbrace = 8 \pi G \gamma \frac{\partial\mathcal{H}_{\mathrm{cl}}}{\partial p_i}
\end{equation}
It is straightforward to check that 
the classical equations of motion lead to the following generalized Friedmann and Raychaudhuri equations
\begin{eqnarray}
    H^2 &=& \frac{8\pi G \rho}{3}+\frac{\sigma^2}{6}, \\\label{Friedmann}
    \dot H &=& -\frac{1}{2}\left( 8\pi G (\rho+P)+ \frac{3\sigma^2}{2}\right).\label{Raychaudhuri}
\end{eqnarray}
Here $\rho = \mathcal{H}_m/v$ and $P = -\partial \mathcal{H}_m/\partial v$ are the energy density and pressure of the matter field respectively, and we will assume the matter is a perfect fluid. If the matter source is a scalar field, the matter part of the Hamiltonian is given by 
\begin{equation}
    \mathcal{H}_\mathrm{m} = \frac{p_{\phi}^2}{2v} + U(\phi)v.
\end{equation}
where $U$ is the corresponding scalar field potential. 
 The conjugate variables are $\phi$ and $p_\phi$ with $p_\phi = \dot{\phi} v$. The fundamental Poisson bracket for these conjugate variables is $\lbrace \phi,p_\phi \rbrace = 1$. 

The mean Hubble rate $H$ can be obtained from the directional Hubble rates as $H=(H_1+H_2+H_3)/3$ where the directional Hubble rates are $H_i=\dot a_i/a_i$. Further, $\sigma^2$ is the shear scalar which measures the anisotropy of the spacetime. The shear scalar is given by
\begin{equation}
    \sigma^2=\frac{1}{3} \left((H_1-H_2)^2+(H_2-H_3)^2+(H_3-H_1)^2 \right).
\end{equation}
For matter with vanishing anisotropic stress, one can show that classically $(H_i-H_j)= \alpha_{ij}/a^3$ where $\alpha_{ij}$ are constants and $a \equiv (a_1 a_2 a_3)^{1/3}$ is the mean scale factor. Thus $\sigma^2$ can also be written as
\begin{equation}\label{eqn:sigma_a_prop}
    \sigma^2= 6\frac{\Sigma^2}{a^6},
\end{equation}
where $\Sigma^2=(\alpha_{12}^2+\alpha_{23}^2+\alpha_{31}^2)/18$  is a constant of motion in the classical theory. 

In this work, we wish to understand possible mechanisms allowing the universe to isotropize. We will consider the case that matter is a perfect fluid with equation of state $w = P/\rho$. Using the generalized Friedmann equation we may find that the energy density and pressure satisfy the following conservation law,
\begin{equation}
    \dot\rho + 3H(\rho + P) = 0.
\end{equation}
This then implies that $\rho \propto a^{-3(1+w)}$ for a constant equation of state. This allows us to intuitively understand why anisotropies are expected to be significant in the very early universe even though our present universe is very nearly isotropic. As shown above, the shear scalar is proportional to $a^{-6}$. Thus, it grows faster than the energy density of ordinary matter fields in a contracting spacetime (equivalently, when evolving backwards in an expanding spacetime). The generalized Friedman equation then implies that the contribution of anisotropies may dominate the dynamics of the early universe. By the same argument, one can see that anisotropies will significantly decrease as the universe expands and becomes macroscopic. Hence, comparing with equation (\ref{eqn:sigma_a_prop}) we see that if our matter field becomes ultra-stiff with $w\approx 1$ during contraction, the energy density will grow similarly to the anisotropic shear and we may expect some isotropization effect if the ultra-stiff matter dominates. Note that in general, the cosmological singularity in  classical Bianchi-I model cosmological is a  cigar-like singularity \cite{jacobs1968spatially}. We now consider the loop quantized effective spacetime of the Bianchi-I model, which is free from singularities.

\subsection{Effective dynamics of Bianchi-I model}
In LQC we apply the methods of LQG to the symmetry reduced cosmological model, in our case the Bianchi-I model. This approach to quantization uses holonomies of the connection and fluxes of the triads as canonical variables. As mentioned above, for many cosmological models it is possible to find a continuum effective description that faithfully approximates the quantum dynamics for a wide variety of quantum states and models. The effective Hamiltonian for the Bianchi-I model for lapse $N=1$ is given by \cite{Ashtekar:2009vc,Chiou:2007sp}\footnote{For an alternate loop quantization of loop quantized Bianchi-I model see \cite{Chiou:2007sp,Martin-Benito:2009xaf}. A recent work shows that this quantization is not viable for both compact as well as non-compact topologies \cite{Motaharfar:2023hil}.},
\begin{equation}
\label{hamiltonian}
    \mathcal{H}_{\mathrm{eff}}^\mathrm{I}=-\frac{v}{8\pi G \gamma^2 \lambda^2} \left(\sin(\bar\mu_1 c_1) \sin(\bar\mu_2 c_2) + \textit{cyclic} \right) + \mathcal{H_\mathrm{m}},
\end{equation}
where $ \lambda^2 = \Delta = 4 \sqrt{3} \pi \gamma \ell_{\mathrm{Pl}}^2 $ is the minimum area eigenvalue in LQG. The edge lengths $\mu_i$ of holonomies are given by 
\begin{equation}\label{mu_term}
    \bar \mu_1=\lambda \sqrt{\frac{p_1}{p_2 p_3}}  \quad    \text{and its cyclic permutations}.
\end{equation}
The numerical simulations for Bianchi-I model performed in the next section use the effective equations of motion obtained form the above Hamiltonian, which are given by
\begin{align}\label{triad_flow}
    \dot{p_1} =&\frac{p_1}{\gamma\lambda}\left(\sin(\overline{\mu}_2c_2) + \sin(\overline{\mu}_3c_3)\right)\cos(\overline{\mu}_1c_1),\\
    \dot{c_1} =& \frac{V}{2\gamma \lambda^2 p_1}\Big[ c_2\overline{\mu}_2\cos(\overline{\mu}_2c_2)\left(\sin(\overline{\mu}_3c_3) + \sin(\overline{\mu}_1c_1)\right) + c_3\overline{\mu}_3\cos(\overline{\mu}_3c_3)\left(\sin(\overline{\mu}_1c_1) + \sin(\overline{\mu}_2c_2)\right) \nonumber\\ 
    & - c_1\overline{\mu}_1\cos(\overline{\mu}_1c_1)\left(\sin(\overline{\mu}_2c_2) + \sin(\overline{\mu}_3c_3)\right) - \left(\sin(\overline{\mu}_1c_1)\sin(\overline{\mu}_2c_2)  + \textit{cyclic} \right) \Big] \nonumber\\ 
    & + 8\pi G\gamma \frac{\partial \mathcal{H}_\mathrm{m}}{\partial p_1},
\end{align}
with similar equations for the other triad and connection components. Unlike classical dynamics, the energy density $\rho$ is bounded in effective spacetime, which can be easily seen by obtaining $\rho$ from the vanishing of the effective Hamiltonian constraint as follows 
\begin{align}
\label{rhomax_I}
    \rho &= \frac{1}{8\pi G\gamma^2\lambda^2}\left(\sin(\overline{\mu}_1c_1)\sin(\overline{\mu}_2c_2) + \textit{cyclic}\right) \leq \rho_\mathrm{max} = \frac{3}{8\pi G\gamma^2\lambda^2} \approx 0.41\rho_\mathrm{pl},
\end{align}
The upper bound $\rho_\mathrm{max}$ is the same as the one obtained in LQC of the isotropic model. Similarly the directional Hubble rates are also bounded:
\begin{equation}
    H_1 = \frac{\dot{a_1}}{a_1} = \frac{1}{2\gamma\lambda}\left(\sin(\overline{\mu}_1c_1 - \overline{\mu}_2c_2) + \sin(\overline{\mu}_1c_1-\overline{\mu}_3c_3) + \sin(\overline{\mu}_2c_2 + \overline{\mu}_3c_3) \right)
\end{equation}
with similar equations for the other directional Hubble rates. Consequently, the shear scalar is also bounded \cite{Gupt:2011jh}: 
\begin{align}
    \sigma^2 =& \frac{1}{3}\left((H_1-H_2)^2 + (H_2-H_3)^2 + (H_3-H_1)^2\right) \label{shear} \\
    =& \frac{1}{3\gamma^2\lambda^2}\Big[\Big(\cos(\overline{\mu}_3c_3)(\sin(\overline{\mu}_1c_1) + \sin(\overline{\mu}_2c_2)) - \cos(\overline{\mu}_2c_2)(\sin(\overline{\mu}_1c_1) + \sin(\overline{\mu}_3c_3))\Big)^2 + \textit{cyclic}  \Big] \label{shearI} \\
    \leq & \sigma_\mathrm{max}^2 = \frac{10.125}{3\gamma^2\lambda^2}\approx 11.57 l_\mathrm{pl}^{-2}. \label{shearmax_I}
\end{align}
Moreover, it has been shown in \cite{Singh:2011gp}, that strong curvature singularities are generically resolved and the spacetime is geodesically complete in the effective Bianchi-I model. The bounce in the Bianchi-I model is often accompanied by Kasner transitions in the geometry of the spacetime, as shown by extensive numerical simulations \cite{Gupt:2012vi}. \textcolor{black}{These numerical simulations show that even though $\rho_{\mathrm{max}}$ and $\sigma^2_{\mathrm{max}}$ are theoretical maxima and due to the joint contribution of energy density and anisotropic shear to the spacetime curvature, the singularity resolution via bounces can occur at values smaller than these upper bounds. }
Note that, unlike classical dynamics, $\sigma^2$ no longer has a simple dependence on the mean scale factor $a$, and the modified generalized Friedmann and Raychaudhuri equations are as yet unknown for the effective dynamics of the Bianchi-I model. In \cite{McNamara:2022dmf}, an interesting  parabolic relation between the energy density and the shear scalar was numerically shown to hold at the quantum bounce.

\subsection{Classical dynamics of Bianchi-IX model}
An important feature of Bianchi-IX models in contrast to Bianchi-I models is that they posses a nonzero intrinsic spatial curvature. In fact, the isotropic FLRW model with $k=1$ is an isotropic limiting case of the Bianchi-IX model, while the Bianchi-I model is a limiting case with zero intrinsic spatial curvature. Compared to the Bianchi-I model, the homogeneous hypersurfaces $\Sigma$ in the case of the Bianchi-IX model have the topology of $\mathbb{S}^3$, possessing the symmetries associated with the three rotations within $\mathbb{S}^3$. The metric can be written as:
\begin{equation}
    \mathrm{d} s^2= - \mathrm{d} t^2 + q_{a b} \mathrm{d} x^a \mathrm{d} x^b = - \mathrm{d} t^2 + \omega^i_a \omega^j_b \delta_{ij} \mathrm{d} x^a \mathrm{d} x^b.
\end{equation}
The physical forms can be written as $\omega ^i_a=a^i (t) ^o\omega^i_a$, where $^o\omega^i_a$ are the fiducial forms given by
\begin{eqnarray}
^o\omega^1_a &=& \sin \beta \sin \gamma (\mathrm{d}\alpha)_a + \cos \gamma (\mathrm{d}\beta)_a, \\
^o\omega^2_a &=& -\sin \beta \cos \gamma (\mathrm{d}\alpha)_a + \sin \gamma (\mathrm{d}\beta)_a, \\
^o\omega^3_a &=& \cos \beta (\mathrm{d}\alpha)_a + (\mathrm{d}\gamma)_a,
\end{eqnarray}
$\alpha$, $\beta$ and $\gamma$ being the angular coordinates on a 3-sphere with a radius $r_o= 2$. The fiducial volume of the hypersurface can be written as $V_0 = l_o^3 = 2 \pi^2 r_o^3$. In terms of fiducial triads ${\mathring e}_{i}^{a}$ and co-triads ${\mathring\omega}_{a}^{i}$, the symmetry reduced Ashtekar variables are given by
\begin{equation}
E_{i}^{a}=\frac{p_i}{l_o}\sqrt{|\mathring q|} {\mathring e}_{i}^{a} \quad \text{and} \quad A_{a}^{i}=\frac{{c}^i}{l_o} {\mathring\omega}_{a}^{i},
\end{equation}
where $|\mathring q|$ is the determinant of the fiducial spatial metric, and the triad variables $p_i$ are given by 
\begin{equation}
p_1 = \mathrm{sgn}(a_1)|a_2 a_3| {l_o}^2, \quad \text{$p_2$ and $p_3$ given by cyclic permutations}, \label{scale factors to triads}.
\end{equation}
The symmetry reduced connections $c^i$ and triads $p_i$ satisfy the Poisson brackets $\{c^i,p_j\} = 8\pi G\gamma \delta^i_j$
as before. Note our description of the triads implies that $a_i = \frac{1}{l_o}\sqrt{\frac{p_jp_j}{p_i}}$ as well.

The classical Hamiltonian for the Bianchi-IX model in terms of symmetry reduced Ashtekar variables is given by ($N=1$) \cite{Wilson-Ewing:2010lkm}:
\begin{eqnarray}
\mathcal{H}_{\mathrm{cl}}^{\mathrm{IX}} &=& -\frac{1}{8 \pi G\gamma^2 \sqrt{|p_1 p_2 p_3|}}\bigg[(p_1 p_2 c_1 c_2 + \textit{cyclic})\nonumber \\
&&  + l_o \epsilon (p_1 p_2 c_3 + \textit{cyclic})+\frac{l_o^2 (1+\gamma^2)}{4}\bigg(2 p_1^2-\frac{p_1^2 p_2^2}{p_3^2} + \textit{cyclic}\bigg)\bigg]+\mathcal{H}_{\mathrm{m}}, \label{C_Hamiltonian}
\end{eqnarray}
where $\epsilon = \pm 1$ depending on whether the triads are right-handed or left-handed respectively. The above Hamiltonian is invariant under reflections in the triad space, thus we choose $\mathrm{sgn}(a_i)=1$ without loss of generality. The equations of motion then come from the Poisson brackets of the variables with the Hamiltonian, and are given by
\begin{eqnarray}
    \dot p_1 &=& \frac{p_i}{\gamma}\left(p_2c_2 +p_3c_3 + l_o\frac{p_2 p_3}{p_1}\right), \\
    \dot c_1 &=& -\frac{1}{\gamma}\left(p_2c_1c_2 +p_3c_1c_3 + l_o(p_2c_2 + p_3c_2) + l_o^2(1+\gamma^2)\left[p_1 + \frac{p_2^2p_3^2}{2p_1^3} - \frac{p_1p_3^2}{2p_2^2} -\frac{p_1p_2^2}{2p_3^2}\right]\right),
\end{eqnarray}
with $\dot p_2$ and $\dot p_3$, and $\dot c_2$ and $\dot c_3$ given by cyclic permutations of above expressions. 
Given these equations of motion, we find our model satisfies a Raychaudhuri equation:
\begin{equation}
    \dot H = -\frac{1}{2}\left(3H^2 + \frac{2\sigma^2}{3} + \frac{32\pi G\rho}{3}\right).
\end{equation} We can also obtain a generalized Friedmann equation 
\begin{equation}
    H^2 = \frac{8\pi G}{3}\rho  + \frac{1}{6}\sigma^2 - \frac{l_o^2}{12}V(p)
\end{equation}
where the spatial curvature potential is given by  
\begin{equation}
V(p) = \frac{1}{p_1p_2p_3}\left[2(p_1^2 + p_2^2 +p_3^2) - \left(\frac{p_1p_2}{p_3}\right)^2 - \left(\frac{p_2p_3}{p_1}\right)^2 - \left(\frac{p_3p_1}{p_2}\right)\right] .
\end{equation}
This geometric potential plays an important role in the mixmaster dynamics on approach to the singularity.

\subsection{Effective Dynamics of Bianchi-IX model}
We now turn to the effective dynamics of the Bianchi-IX model in loop quantum cosmology. For models with both anisotropy and spatial curvature, holonomies over open edges have to be considered for loop quantization because expressing the field strength operator using only holonomies of closed loops does not lead to an algebra of almost periodic functions as needed. The quantization ambiguities in this procedure lead to two different quantizations which exist in the literature. There is the `A' quantization \cite{Wilson-Ewing:2010lkm}, which considers the holonomies of the Ashtekar-Barbero connection to represent the curvature. Alternatively, in the `K' quantization the extrinsic curvature, rather than the Ashtekar-Barbero connection, is used as the variable conjugate to the triads to construct the curvature \cite{Singh:2013ava}. The `K' quantization of the Bianchi-IX spacetimes is consistent with the Bianchi-I quantization, whereas the `A' quantization does not obtain the above Bianchi-I effective theory as a limiting case (note that the Bianchi-I quantization does not suffer from such an ambiguity). As this work compares results for Bianchi-IX and Bianchi-I spacetimes, it is natural to consider the `K' quantization.\footnote{Similar to  results for Bianchi-I models, quantizing Bianchi-IX spacetimes via either method results in singularity resolution. However, `A' quantization requires inverse triad corrections to obtain a generic singularity resolution \cite{Gupt:2011jh} while this is not the case for `K' quantization.} 
The `K' quantization for Bianchi-IX spacetimes has been obtained in \cite{Singh:2013ava}, where the effective Hamiltonian for $N=1$ was found to be
\begin{equation}
\mathcal{H}_{\mathrm{eff}}^{\mathrm{IX}} = -\frac{\sqrt{p_1 p_2 p_3}}{8 \pi G\gamma^2} \bigg[ \frac{\sin(\bar\mu_1 \gamma k_1)\sin(\bar\mu_2 \gamma k_2)}{\lambda^2} +\frac{l_o^2}{4p_1 p_2 p_3}\bigg(2 p_1^2-\frac{p_1^2 p_2^2}{p_3^2} \bigg) + \textit{cyclic} \bigg] +\mathcal{H}_{\mathrm{m}}. \label{Hamiltonian}
\end{equation}
where the extrinsic curvature is given by $K^{i}_a=k^{i} (^o \omega^i_a) / l_o$, and the standard Poisson brackets are 
\begin{equation}
    \lbrace k^i,p_j \rbrace = 8\pi G \delta^i_j.
\end{equation}

The resulting effective Hamilton's equations for $p_1$ and $k_1$ are:
\begin{eqnarray}
\dot p_1 &=& -8\pi G \frac{\partial\mathcal{H}_{\mathrm{eff}}^{\mathrm{IX}}}{\partial k_1} = \frac{p_1}{\gamma \lambda} (\sin(\bar\mu_2 \gamma k_2)+\sin(\bar\mu_3 \gamma k_3))\cos(\bar\mu_1 \gamma k_1), \label{p1dot} \\
\dot k_1 &=& 8\pi G \frac{\partial\mathcal{H}_{\mathrm{eff}}^{\mathrm{IX}}}{\partial p_1} \nonumber, \\
&=& -\frac{\lambda}{2\gamma^2}\frac{1}{\bar\mu_1}\bigg[\frac{1}{\lambda^2}(\sin(\bar\mu_1 \gamma k_1)\sin(\bar\mu_2 \gamma k_2) + \textit{cyclic})\nonumber \\
&& +\frac{l_o^2}{4p_1 p_2 p_3}\bigg(6 p_1^2-2p_2^2-2p_3^2-3\frac{p_1^2 p_2^2}{p_3^2}+5\frac{p_2^2 p_3^2}{p_1^2}-3\frac{p_3^2 p_1^2}{p_2^2} \bigg) \nonumber \\
&& +\frac{\gamma}{\lambda^2}\bigg(k_1 \bar\mu_1 \cos(\bar\mu_1 \gamma k_1)(\sin(\bar\mu_2 \gamma k_2)+\sin(\bar\mu_3 \gamma k_3)) \nonumber \\
&& -k_2 \bar\mu_2 \cos(\bar\mu_2 \gamma k_2)(\sin(\bar\mu_1 \gamma k_1)+\sin(\bar\mu_3 \gamma k_3)) -k_3 \bar\mu_3 \cos(\bar\mu_3 \gamma k_3)(\sin(\bar\mu_2 \gamma k_2)+\sin(\bar\mu_1 \gamma k_1)\bigg)\bigg]\nonumber \\
&& + 8\pi G \frac{\partial \mathcal{H}_\mathrm{m}}{\partial p_1,}
\end{eqnarray}
where the dynamical equations for the other variables can be obtained from cyclic permutations of the above equations. 

From vanishing of the Hamiltonian constraint, the energy density can be expressed as
\begin{eqnarray}
\rho &=& \frac{1}{8 \pi G\gamma^2}\bigg[\frac{\sin(\bar\mu_1 \gamma k_1)\sin(\bar\mu_2 \gamma k_2)}{\lambda^2}
+\frac{l_o^2}{4p_1 p_2 p_3}\bigg(2 p_1^2-\frac{p_1^2 p_2^2}{p_3^2} \bigg) + \textit{cyclic})\bigg].
\end{eqnarray}
Like in the Bianchi-I case, it is clear that the first term is bounded. While the second term being bounded is unclear from the equation itself, it may be argued that the $p_\mathrm{i}$ and their inverses always are bounded between zero and infinity. Hence, the energy density is bounded \cite{Saini:2017ggt}. 
While we again do not have analogs of the modified generalized Friedmann and Raychaudhuri equations in the quantized setting, we find that the  directional Hubble rate $H_1$ is given by,
\begin{equation}
H_1 = \frac{1}{2 \gamma \lambda} (\sin(\bar\mu_1 \gamma k_1 -\bar\mu_2 \gamma k_2) + \sin(\bar\mu_1 \gamma k_1 -\bar\mu_3 \gamma k_3)+\sin(\bar\mu_2 \gamma k_2 + \bar\mu_3 \gamma k_3) ),
\end{equation}
and $H_2, H_3$ can be obtained by cyclic permutations. Note that the Hubble rates are bounded as well.
Similarly, we find that the anisotropic shear is bounded, as seen in the following expression \cite{Gupt:2011jh}
\begin{align}
    \sigma^2 =& \frac{1}{3\gamma^2 \lambda}\left[\sin{(\bar\mu_1\gamma k_1 + \bar\mu_2\gamma k_2)} + \sin(\bar\mu_2\gamma k_2 + \bar\mu_3\gamma k_3) + \sin(\bar\mu_3 \gamma k_3 + \bar\mu_1\gamma k_1) \right]^2 + \textit{cyclic}\\
    \leq& \frac{10.125}{3\gamma^2 \lambda}\approx 11.57 l_\mathrm{pl}^{-2}.\label{shearmax_IX}
\end{align}
Note that this is the exact same bound as found in the Bianchi-I effective spacetime in equation (\ref{shearmax_I}), as this expression may be derived from the directional Hubble rates using equation \eqref{shear}. \textcolor{black}{As in the case of Bianchi-I model in LQC, the bounce in Bianchi-IX spacetime can occur at energy density and anisotropic shear at values below their maximum allowed values. }\\

In the following sections, we perform the numerical simulations of effective dynamics for Bianchi-I and Bianchi-IX spacetimes. 
The initial conditions are chosen in the contracting branch far from the bounce where the universe is macroscopic and the effective spacetime is well-approximated by classical dynamics. Apart from the matter content, we have six phase space variables in both Bianchi-I and Bianchi-IX models, and one constraint; namely the Hamiltonian constraint. \textcolor{black}{For simulations, we first choose the initial values of $c_1, c_2, p_1, p_2, p_3, \phi$ and $\rho$. The initial conditions chosen for all the simulations in the manuscript correspond to a  classical macroscopic universe with energy density positive. The value of connection component $c_3$ is determined using the vanishing of Hamiltonian constraint. The values of $c_1$ and $c_2$ are chosen in range $c_1, c_2 \in [-0.25,0]$. The values of triads are chosen in range $p_1,p_2,p_3 \in [10000,30000]$, and the value of initial energy density is chosen in the range $\rho \in [0,10^{-4}]$. The value of scalar field was chosen in the range $\phi \in [0,0.4]$.}
\textcolor{black}{
The initial value of $p_\phi$ can be obtained from the initial values of $\phi$ and $\rho$.\footnote{\textcolor{black}{The actual allowed range of $\phi$ varies with the potential parameters. As an example, for the ekpyrotic potential \eqref{Ekpyrotic_potential} with $U_o = 0.25$, $\sigma_1 = 0.3 \sqrt{8 \pi}$ and $\sigma_2 = 0.09 \sqrt{8 \pi}$, $\phi$ can not be more than 0.0007 for $\rho = 10^{-4}$. Using the relationship between $p_\phi$ and $\dot \phi$, the range of $\dot \phi$ falls roughly between 
$\dot \phi \in [0,1.5 \times 10^{-2}]$ for the ekpyrotic potential and $\dot \phi \in [0,1]$ for the ekpyrotic-like potential.}}} The effective equations of motion for the triad and connection variables described earlier are then used to obtain the time evolution. For both Bianchi-I and Bianchi-IX models, $500$ simulations each are performed for both ekpyrotic and ekpyrotic-like potentials as well as for the massless scalar field for comparison. When compared to the massless case, the values of $c_1$, $c_2$, all triads and $\phi$ and $\rho$ were taken as same.  \textcolor{black}{As discussed earlier in the Introduction, recall that we are considering a macroscopic contracting universe in which the ekpyrotic field is initially in the positive regime moving towards smaller $\phi$ values. While ekpyrosis can occur when the field enters the negative well, the initial conditions considered in our work do not assume existence of ekpyrosis.} \\

\textcolor{black}{Note that the purpose of this work is not to obtain viable ekpyrotic cosmology per se, rather it is to study the effect of the ekpyrotic potential on anisotropies starting from generic initial conditions corresponding to macroscopic contracting universe. Further, the initial conditions are not specifically tailored to obtain a definite phase of ekpyrosis (a phase with $w \gg 1$), instead the existence and duration of the ekpyrosis is left to be determined by the dynamics while setting random initial conditions for the scalar field from a chosen range of values corresponding to a macroscopic contracting universe. In other words, an extended phase of ekpyrosis is not guaranteed by the initial conditions chosen in our numerical simulations. 
We have chosen positive initial values for the scalar field in this study, wherein the permitted initial range of values of $\phi$ and $\dot \phi$ are determined by the fact that we set our initial conditions in the far past where the universe is large and contracting and the total energy density is very small compared to the Planck scale. Further, as noted above, in LQC  there are also global bounds on energy density and shear scalar determined by the underlying quantum geometry. It is also important to note that at the bounce these theoretical maximum values are generally not saturated and singularity resolution can occur in anisotropic spacetime at values lower than $\rho_{\mathrm{max}}$ and $\sigma^2_{\mathrm{max}}$. This produces a constraint on the allowed initial conditions by the Hamiltonian constraint in loop quantum cosmology. As an example, it is not possible to consider the initial value of $\phi$ of the order unity and large $\dot \phi$ in our numerical simulations for the ekpyrotic potential. For concreteness, consider the case of the ekpyrotic potential \eqref{Ekpyrotic_potential} with $U_o = 0.25$, $\sigma_1 = 0.3 \sqrt{8 \pi}$ and $\sigma_2 = 0.09 \sqrt{8 \pi}$ in Bianchi-I spacetime as an example.} If one takes $\phi \approx 1$ as the initial condition, then $U \approx 0.1$ in Planck units. Even if the kinetic energy of the field is vanishing, this contribution from potential energy means that the universe already is in the Planck regime in terms of the value of energy density. Thus, such an initial condition is not compatible with an initial macroscopic classical universe with a small energy density. Since the maximum bound on energy density is 0.41, this means that the maximum allowed contribution from kinetic energy term is 0.31 in Planck units. Even if one assumes that energy density at the bounce reaches the maximum theoretical value, one finds that the magnitude of $\dot \phi$ is approximately 0.8. Therefore, one can not have initial conditions with $\phi$ of the order unity and $\dot \phi$ very large in Planck units in LQC. In contrast, in case of the ekpyrotic-like potential which is negative definite, it is indeed possible to consider large and positive initial values of $\phi$. \textcolor{black}{However, the ekpyrotic-like potential already assumes very small values in the range of initial values considered in this manuscript, i.e. $\phi \in [0,0.4]$. The ekpyrotic-like potential keeps decreasing in magnitude gradually for higher values of $\phi$, such that going to values higher than the considered range is unlikely to yield radically different results than those obtained in this manuscript. 
 Note that both the potentials have a single global minimum adjacant to $\phi=0$. While the ekpyrotic-like potential remains negative and approaches vanishing values on either side of the minimum, the ekpyrotic potential becomes positive on the right branch of the minima beyond $\phi=0$. Our initial conditions for the field are set on the right edge of the potential minimum with the field evolving towards the minimum as the universe contracts. While the initial conditions considered in this manuscript result in 
short-lived phases of ekpyrosis, we still find significant isotropization.} \\

\section{Ekpyrosis in Bianchi-I spacetime in LQC}\label{sec:BianchiI_potential}

We now consider the effective dynamics of Bianchi-I spacetimes with both ekpyrotic and ekpyrotic-like potentials. In particular, we are interested in determining the effect of these potentials on causing isotropization of the universe before and after the bounce. To investigate this, we will compare the evolution of universes under these potentials to that of universes with only a massless scalar field. We are able to study the direct effect of the potential on a given simulation by choosing initial conditions in the contracting semiclassical regime and then evolving from these initial conditions both with and without the potential. In particular, we study isotropization at the last bounce before entering a large universe phase as we expect this bounce to determine the semiclassical evolution of the universe.

\subsection{Ekpyrotic potential}\label{subsec:BianchiI_Cycle}

\begin{figure}
\centering
\begin{subfigure}
    \centering
    \includegraphics[width=0.49\linewidth]
    {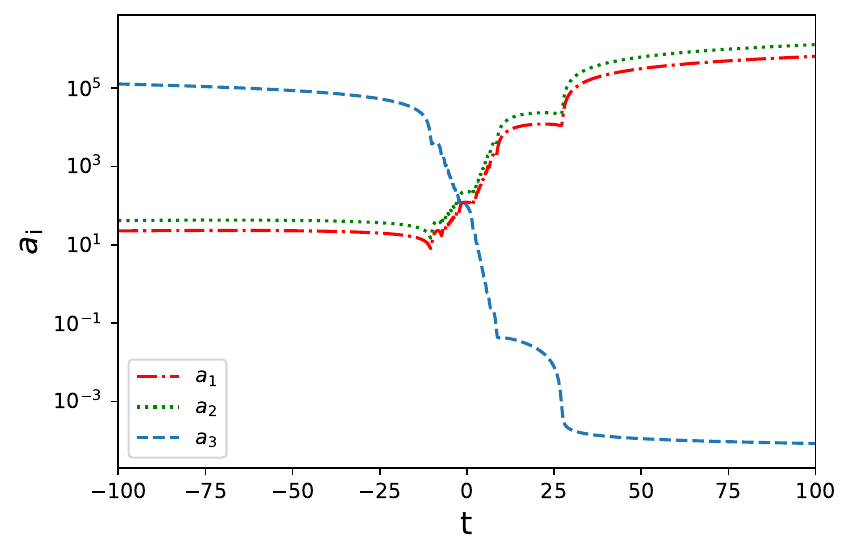}
\end{subfigure}
\begin{subfigure}
    \centering
    \includegraphics[width=0.49\linewidth]
    {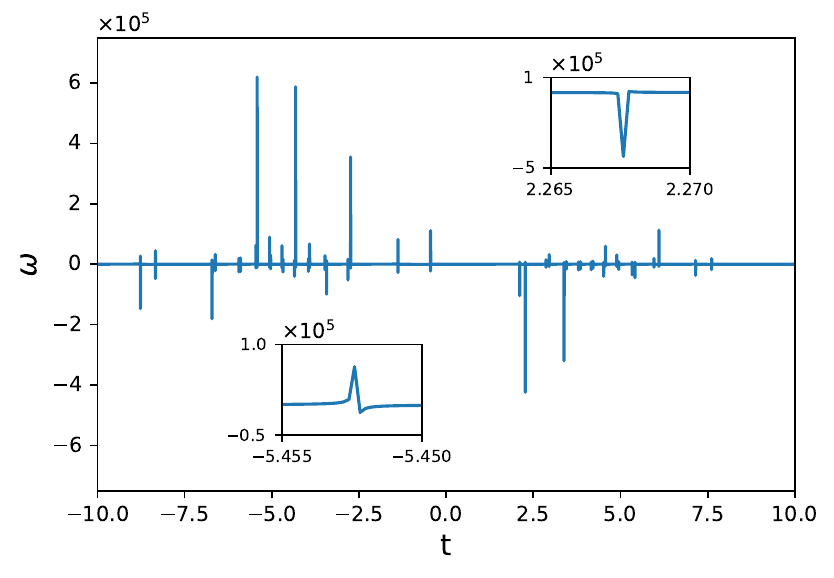}
\end{subfigure}
\begin{subfigure}
    \centering
    \includegraphics[width=0.49\linewidth]
    {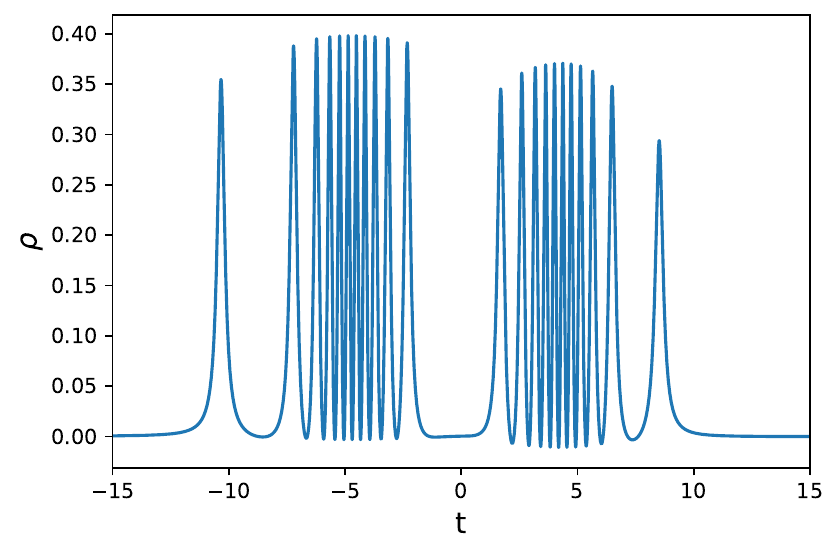}
\end{subfigure}
\begin{subfigure}
    \centering
    \includegraphics[width=0.48\linewidth]
    {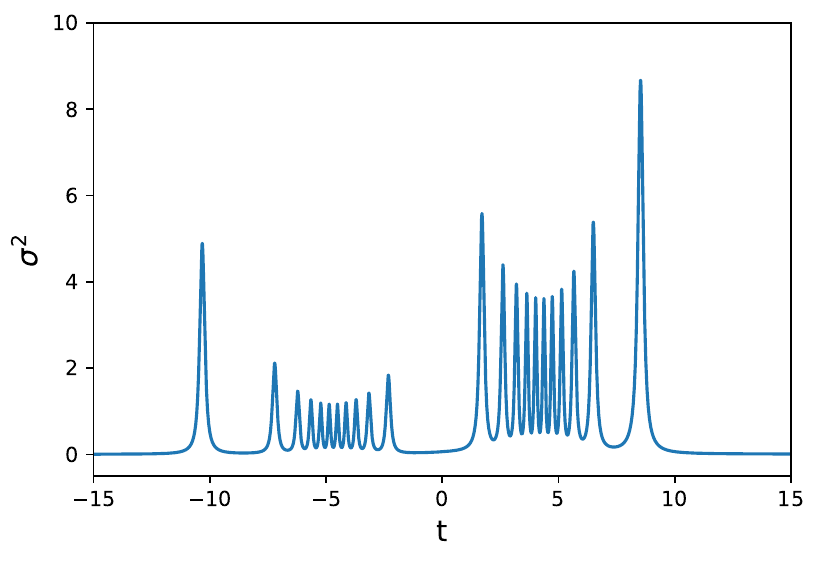}
\end{subfigure}
\caption{An example of the non-singular evolution of directional scale factors, equation of state, energy density and anisotropic shear in the bounce regime for effective Bianchi-I spacetime coupled to the ekpyrotic scalar field with $U_0 = 0.25$ in \eqref{Ekpyrotic_potential}. \textcolor{black}{If one considers time from negative to positive three scale factors are
decreasing before $t = 0$ and then two are increasing while one is decreasing after the bounce. The universe has a point like approach to singularity which after resolution due to quantum geometry effects yields a cigar like evolution as is generally the case for Kasner transitions in LQC \cite{Gupt:2012vi}.} It is typical to get multiple bounces before the universe eventually enters a macroscopic expanding regime. Unlike the classical theory, the scale factors have finite nonzero values throughout the evolution and the energy density and anisotropic shear do not diverge. These features are seen in all simulations discussed in this manuscript.}
\label{fig:Bianchi-I_Cycle_0.25_evolution}
\end{figure}

\begin{figure*}
\centering
\begin{subfigure}
    \centering
    \includegraphics[width=0.49\linewidth]{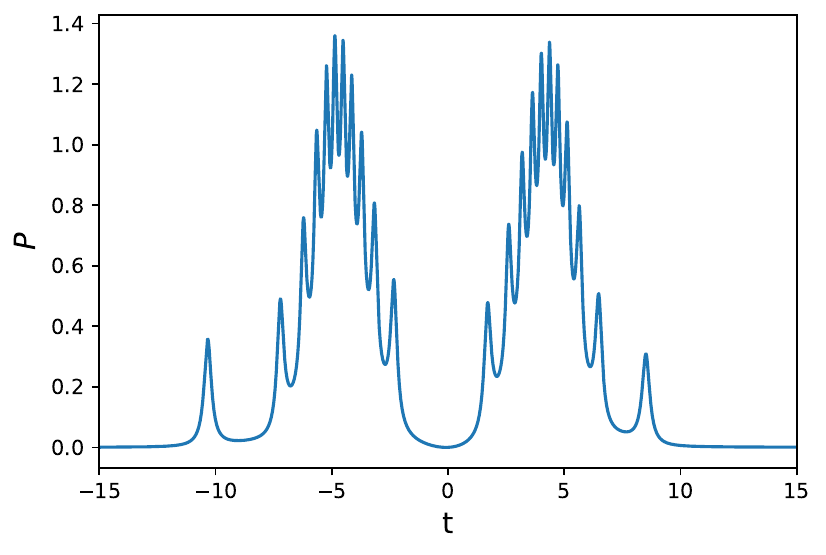}
\end{subfigure}
\begin{subfigure}
    \centering
    \includegraphics[width=0.48\linewidth]{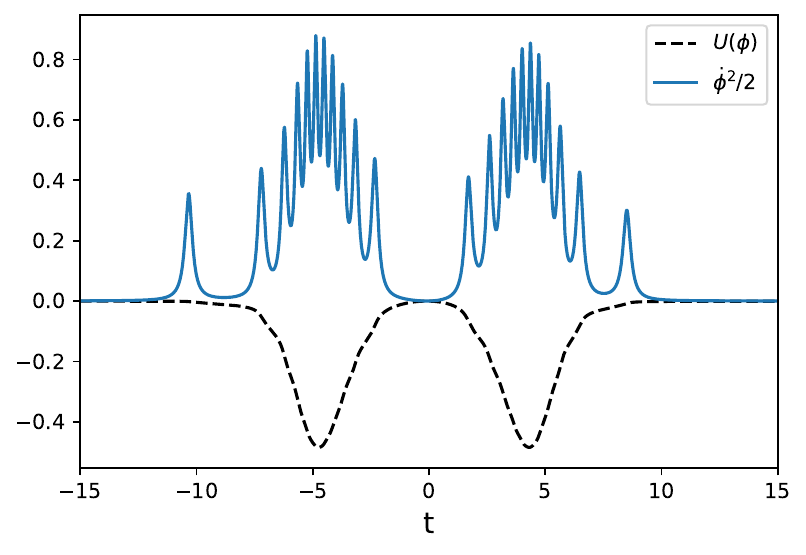}
\end{subfigure}
\caption{The time evolution of the pressure and potential and kinetic energies corresponding to the example shown in Figure 1. In the case of the ekpyrotic scalar field, the pressure remains positive definite. }
\label{fig:BianchiI_Cycle_0.25_Energy}
\end{figure*}

We now consider Bianchi-I spacetimes with the ekpyrotic potential 
and compare this with the massless scalar field in order to determine  the way the potential affects anisotropies at the bounce. 
The potential for the ekpyrotic field is given by
\begin{equation}\label{Ekpyrotic_potential}
    U_{\mathrm{E}}(\phi) = U_0\left(1-e^{-\sigma_1\phi}\right)\exp\left({-e^{-\sigma_2\phi}}\right)
\end{equation}
where $U_0$, $\sigma_1$ and $\sigma_2$ are parameters of the potential.
Using the effective equations of motion laid out in the previous section, we perform simulations of the dynamics in this case for a variety of initial conditions, for both the massless scalar field and the ekpyrotic scalar field. The details of picking the initial conditions were discussed earlier. Throughout these simulations we take the potential parameters aside from $U_0$ to be $\sigma_1 = 0.3\sqrt{8\pi}$ and $\sigma_2 = 0.09\sqrt{8\pi}$. For $U_0$, we pick two different values, $U_0=0.25$ and $U_0 = 1.25$, in order to compare how scaling the potential affects ekpyrosis. In Fig. \ref{fig:Bianchi-I_Cycle_0.25_evolution}, we show the typical evolution of various quantities in the case of an ekpyrotic field in effective Bianchi-I universe as exemplified by a simulation with initial conditions $c_1=-0.22031, c_2=-0.10265, p_1=12281, p_2=27618, p_3=10287, \phi=4 \times 10^{-5}, \rho=6.05 \times 10^{-5} $ and $c_3$ solved from the Hamiltonian constraint to be $c_3 = -5.1228$. As seen in Fig. \ref{fig:Bianchi-I_Cycle_0.25_evolution}, the classical singularity is resolved by a quantum bounce in the effective Bianchi-I spacetime and the scale factors begin to approach their classical behavior when the universe emerges from the bounce regime and starts expanding again. Due to the interplay between the energy density and anisotropies in the bounce regime, one typically obtains multiple bounces for generic initial conditions, after which the universe enters an expanding phase reaching macroscopic volumes. As seen in the equation of state plot, the ekpyrotic field behaves as normal matter ($w < 1$) for most of its evolution and ekpyrosis ($w > 1$) occurs only in the bounce regime. Interestingly, the bounce regime is dotted with multiple short spurts of ekpyrosis rather than having a single long phase of ekpyrosis. This behavior is consistent with earlier studies using the ekpyrotic field \cite{Li:2020pww}. Interestingly, the energy density almost attains its maximum value at all the bounces while this is not the case for the shear scalar in Fig. \ref{fig:Bianchi-I_Cycle_0.25_evolution}. This can be seen as an effect of the ekpyrotic field for which the equation of state can become ultra-stiff in the bounce regime as seen in the equation of state plot, which leads to energy density increasing much faster than shear scalar in the bounce regime. Thus, the energy density very quickly approaches its maximum value near the bounce as will be borne out by our simulations discussed below. As shown in Fig. \ref{fig:Bianchi-I_Cycle_0.25_evolution} and all simulations below, a general feature of these spacetimes equipped with an ekpyrotic field is an equation of state that is very large and negative in the bounce regime. This can be explained by regions where the ekpyrotic potential becomes negative in Fig. \ref{fig:BianchiI_Cycle_0.25_Energy} while the kinetic energy remains positive. This results in a range 
 of very small and negative values of the energy density. Since the equation of state goes as $w = P/\rho$ and the pressure remains positive definite, this leads to equation of state becoming negative.

We now compare the behavior of the energy density and anisotropic shear at the bounce for the effective Bianchi-I universe with and without the ekpyrotic potential. This is explored in Figures~\ref{fig:BianchiI_Cycle_0.25} ~and~\ref{fig:BianchiI_Cycle_1.25} for $U_0=0.25$ and $1.25$ respectively. As discussed above, the ekpyrotic (and ekpyrotic-like) potential is of interest for reducing the degree of anisotropy due to its behavior allowing an equation of state with $w>1$. This then may allow for the energy density to dominate over the anisotropy during the bounce regime. If the ekpyrotic potential causes isotropization at the bounce, we expect the energy density to increase in comparison to the massless scalar case, and the anisotropic shear to decrease. Further, a larger value of $U_0$ may be expected to result in a more isotropic bounce, thus we have considered two different values of $U_0$. In all plots, the values of the energy density and shear scalar are divided by the maximum value of these quantities in the effective dynamics of Bianchi-I given in equations \eqref{rhomax_I} and \eqref{shearmax_I} respectively. In addition we also plot the ratio 
\begin{equation}\label{R_sigma}
R_\sigma=\frac{\sigma_\mathrm{b}^2/\sigma_\mathrm{max}^2}{\sigma_\mathrm{b}^2/\sigma_\mathrm{max}^2 + \rho_\mathrm{b}/\rho_\mathrm{max}}
\end{equation}
which captures the relative strength of anisotropic shear and energy density at the last bounce. 
Since, compared to the massless scalar field, the ekpyrotic field in general leads to multiple bounces, we have chosen the simulation length such that in most cases the universe eventually approaches a classical regime in the expanding branch. It is important to note that when we refer to `last bounce' in this manuscript, it refers to the last bounce in the chosen time range of the simulation only. In these figures, the diagonal is also plotted for ease of viewing the location of points.

\subsubsection{Effective Bianchi-I dynamics with ekpyrotic potential: $U_0=0.25$}

\begin{figure}
\centering
\begin{subfigure}
       \centering
   \includegraphics[width=0.49\linewidth]
    {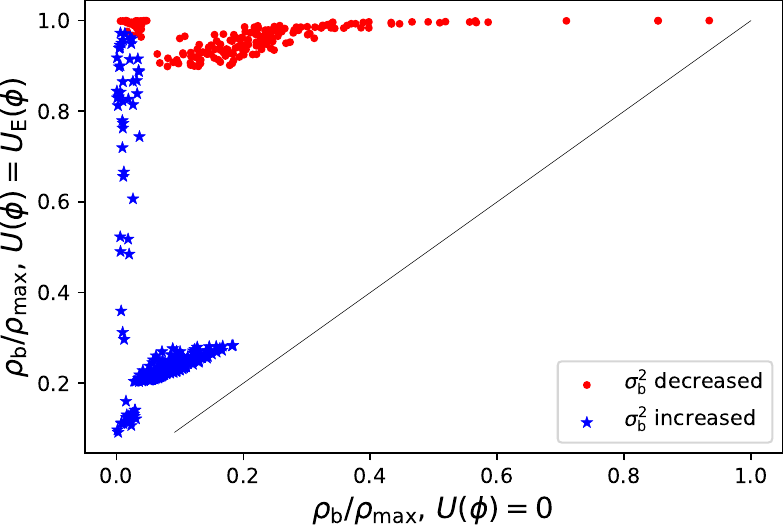}
\end{subfigure}
\begin{subfigure}
   \centering
   \includegraphics[width=0.49\linewidth]
   {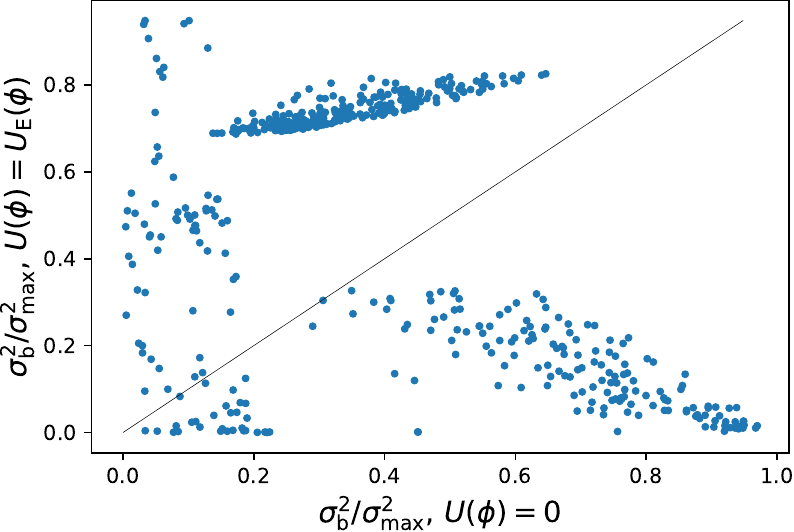}
\end{subfigure}
\begin{subfigure}
   \centering
   \includegraphics[width=0.49\linewidth]
   {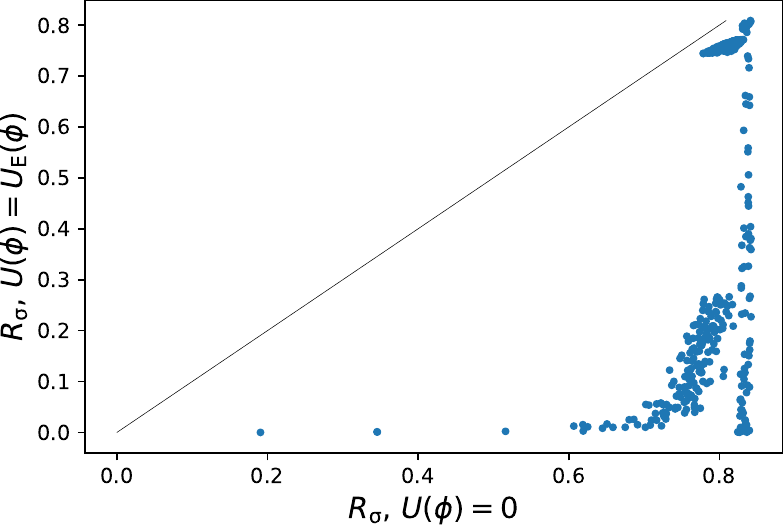}
\end{subfigure}
\caption{Bianchi-I effective spacetime with an ekpyrotic field having $U_0 = 0.25$: comparison of energy density $\rho_\mathrm{b}/\rho_\mathrm{max}$, shear $\sigma_\mathrm{b}^2/\sigma_\mathrm{max}^2$, and the relative strength of anisotropies $R_\sigma$ at the last bounce in case of the ekpyrotic potential (y-axis) versus the massless scalar case (x-axis) for 500 simulations with matching initial conditions. The dots represent one simulation each, and the diagonal line $(x=y)$ is shown for ease of comparison. The dots in the first and second panel are color coded to provide additional information. The same color coding is used in all comparison plots in this manuscript. The blue dots in the energy density plot represent simulations where the shear scalar at the bounce increased relative to the massless case. Similarly, the blue dots in shear scalar plot represent simulations where energy density at the bounce was higher relative to the corresponding massless case. The red dots represent simulation with the opposite trend. We see here that the energy density at the bounce increased compared to the massless case in all the simulations whereas the shear is not necessarily decreased in all the simulations. Nevertheless, the $R_\sigma$ plot shows that the relative strength of anisotropies decreased in all the simulations.}
\label{fig:BianchiI_Cycle_0.25}
\end{figure}

We first discuss the case for $U_0=0.25$, which is shown in Fig. \ref{fig:BianchiI_Cycle_0.25}. The first panel of Fig. \ref{fig:BianchiI_Cycle_0.25} shows $\rho_\mathrm{b} /\rho_\mathrm{max}$ at the last bounce for the case of ekpyrotic potential in comparison to its value at the bounce for the case of a massless scalar field for matched simulations, i.e. those which share the same initial conditions for $p_1, p_2, p_3, c_1, c_2, \rho $ and $\phi$ ($c_3$ is solved for in each case to ensure that the Hamiltonian vanishes, but since we choose semiclassical initial conditions where the potential energy is much smaller than the kinetic energy the initial values obtained for $c_3$ with and without potential are very close). Each point in the plot corresponds to one particular simulation. There are several noticeable features in the energy density plot. First of all, we note that all the dots, representing one simulation each, are above the diagonal line. Thus compared to the massless case, the ekpyrotic potential leads to an increase in energy density at the bounce for all the simulations. Further, the plots in Fig. \ref{fig:BianchiI_Cycle_0.25} are color coded to provide additional information about the simulations represented by the dots. The blue dots in the energy density plot represent those simulations for which the shear scalar at the bounce increased compared to the massless case, while the shear scalar at the bounce decreased compared to the massless case for the red dots. Interestingly, we find that the blue and red dots are separated from each other in the energy density plot. However, this feature is not universal, and a sharp boundary separating the red and blue dots in the energy density plots is only seen in the case of Bianchi-I effective spacetime. As we shall see in the next section, this does not hold true in the effective dynamics of Bianchi-IX model. Further,  we notice yet another type of clustering in the location of dots in the energy density plots in Fig. \ref{fig:BianchiI_Cycle_0.25}. There are two clusters - the ones lying near the top of the plot for which the energy density at the bounce is sharply increased compared to the massless case, while the other cluster lying close to the diagonal line corresponds to a modest rise in energy density at the bounce compared to the massless case. As we shall show below, this clustering disappears when the strength of the potential is increased and all the dots move up, showing a sharp increase in energy density at the bounce compared to the massless case.

The second panel of Fig. \ref{fig:BianchiI_Cycle_0.25} shows the value of the shear scalar $\sigma_\mathrm{b}^2/\sigma_\mathrm{max}^2$ at the last bounce in the time range of all simulations compared to the corresponding massless case with matched initial conditions. In sharp contrast to the energy density plot, we find that the shear scalar does not always decreased when compared to the massless scalar case. 
Some clustering of the points is seen, but this is not a universal feature and disappears when the strength of the potential is increased, as will be seen below. The plot of the shear scalar shows that an increase in the energy density at the bounce (compared to the massless case) is not always accompanied by a decrease in the shear scalar. This appears to suggest that the anisotropies are not necessarily dampened by the ekpyrotic potential. 
While the ekpyrotic potential directly affects the energy density, it has only an indirect effect on the shear scalar. Thus, we should consider a better indicator of the effect of the ekpyrotic potential on the significance of anisotropies in the bounce regime.

The third panel of Fig. \ref{fig:BianchiI_Cycle_0.25} shows the value of the ratio $R_\sigma$ from \eqref{R_sigma} at the last bounce in the time range of simulations for the ekpyrotic field versus the massless scalar field. As all points are below the diagonal line, this shows that the relative magnitude of the anisotropies at the bounce is decreased in all the simulations. As we expect isotropization to depend upon the relative magnitude of the isotropic energy density and the anisotropic shear at the bounce, we observe that isotropization is achieved by the presence of the ekpyrotic field in comparison to the massless scalar field.

Before moving on to the case with an increased strength of the potential, we discuss the possible reasons for the increase in the shear scalar at the bounce compared to the massless case as seen in many simulations in Fig. \ref{fig:BianchiI_Cycle_0.25}, despite the occurrence of ekpyrosis in the bounce regime. Specifically, while the energy density increased in $100\%$ of the simulations, we find that only $37.60\%$  of all simulations had $\sigma_{U_
\mathrm{E}(\phi)}^2 \leq \sigma_{0}^2$ at the bounce where $\sigma_{U_E(\phi)}^2$ is the shear scalar with the potential, and $\sigma_0^2$ is the shear scalar for the massless scalar field. In the previous work \cite{McNamara:2022dmf}, it was shown that there exists an inverted parabolic relation between the shear scalar and the energy density at the bounce in the effective Bianchi-I model, which can be used to understand the above-mentioned counter-intuitive behavior of the shear scalar. In particular, due to this relation between the shear scalar and energy density, increasing the energy density from a low value is initially accompanied by an increase in the shear scalar. This behavior is observed until the shear scalar meets its maximum value, peaking when the increased energy density reaches roughly half the maximum value $\rho_\mathrm{max}$. Only when the energy density is increased well beyond the halfway mark do we see a decrease in the shear scalar.

\begin{figure}
\centering
\begin{subfigure}
       \centering
   \includegraphics[width=0.49\linewidth]
   {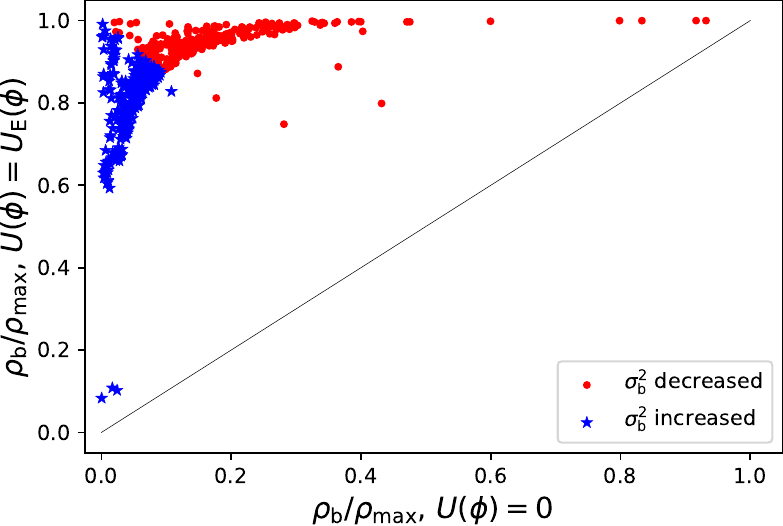}
\end{subfigure}
\begin{subfigure}
   \centering
   \includegraphics[width=0.49\linewidth]
   {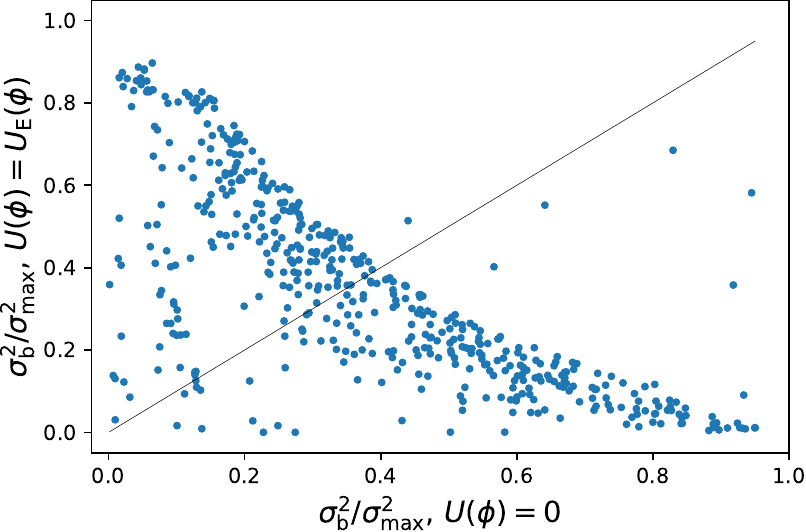}
\end{subfigure}
\begin{subfigure}
   \centering
   \includegraphics[width=0.49\linewidth]
   {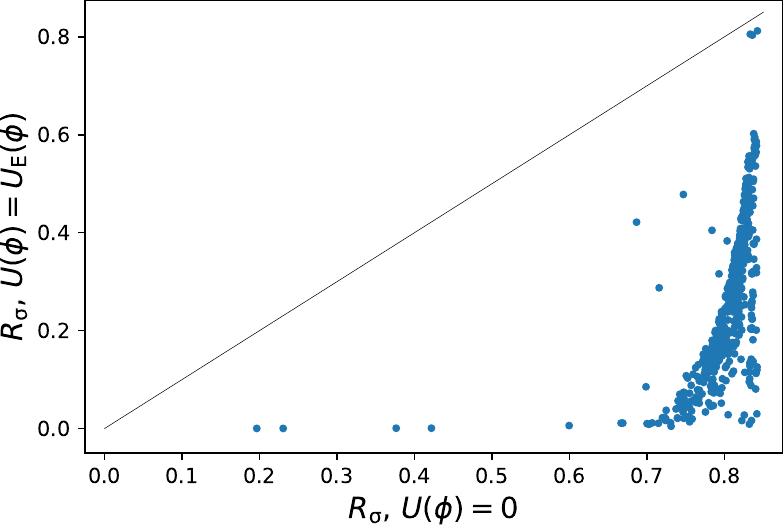}
\end{subfigure}

\caption{Bianchi-I effective spacetime with an ekpyrotic field ($U_0 = 1.25$): comparison of energy density $\rho_\mathrm{b}/\rho_\mathrm{max}$, shear $\sigma_\mathrm{b}^2/\sigma_\mathrm{max}^2$, and the relative strength of anisotropies $R_\sigma$ at the last bounce between the ekpyrotic potential (y-axis) and the massless scalar case (x-axis) with matching initial conditions. Each plot contains 500 dots representing one simulation each. The color coding scheme used for dots in the energy density and shear scalar plots is the same as in the previous figure.}

\label{fig:BianchiI_Cycle_1.25}
\end{figure}

\subsubsection{Effective Bianchi-I dynamics with ekpyrotic potential: $U_0=1.25$}

Having studied the isotropization effects of the ekpyrotic model, we now study the effects of increasing the strength of the potential.  Fig. \ref{fig:BianchiI_Cycle_1.25} summarizes the effects of the ekpyrotic potential with $U_0 = 1.25$ in comparison to the massless scalar field with matched initial conditions as described earlier. As expected, the arguments presented above regarding the isotropization effect of the ekpyrotic field in the case of $U_0 = 0.25$ broadly apply to this case as well, as can be seen in Fig. \ref{fig:BianchiI_Cycle_1.25}. However, we are now interested in comparing the results for $U_0 = 0.25$ and $U_0 = 1.25$ shown in Fig. \ref{fig:BianchiI_Cycle_0.25} and Fig. \ref{fig:BianchiI_Cycle_1.25}. Comparing the energy density plots in the two cases, we note that as the strength of the potential is increased, the cluster of simulations at $U_0=0.25$ in Fig. \ref{fig:BianchiI_Cycle_0.25} where the energy density at the bounce is only marginally increased (i.e. the cluster near the diagonal line) has moved up substantially and merged with the points near maximum energy density at $U_0 = 1.25$ in Fig. \ref{fig:BianchiI_Cycle_1.25}. This shows that increasing the strength of the ekpyrotic potential leads to a higher probability of obtaining a substantial increase in the energy density at the bounce, and in turn leads to higher isotropization at the bounce. Similarly, the clusters in the shear plots at $U_0=0.25$ shown in Fig. \ref{fig:BianchiI_Cycle_0.25} have disappeared at $U_0=1.25$ as shown in Fig. \ref{fig:BianchiI_Cycle_1.25} when the strength of the potential is increased. In particular, in the third panel of Fig. \ref{fig:BianchiI_Cycle_0.25} where the relative importance of anisotropies at the bounce is plotted, there is a cluster of dots near the diagonal line representing only a marginal decrease for a large subset of simulations. In comparison, in the corresponding plot in Fig. \ref{fig:BianchiI_Cycle_1.25}, almost all points are substantially below the diagonal line, indicating strong isotropization in almost all simulations.

The above analysis using extensive numerical simulations establishes that the ekpyrotic potential leads to isotropization at the bounce in the case of an effective Bianchi-I spacetime. In particular, the ekpyrotic field often results in an increased energy density at the bounce, while the effect on the anisotropic shear is less consistent. This is due to the fact that having ekpyrosis in the bounce regime with $w>1$ contributes directly to an increase in the energy density while the effect on the shear scalar is indirect through the equations of motion. Nevertheless, the relative strength of the anisotropies at the bounce is decreased in all cases, leading to an increased isotropization. Further, the extent of this isotropization effect depends on the strength of the ekpyrotic potential. In particular, the stronger ekpyrotic potential with $U_0=1.25$ leads to a substantial decrease in the relative strength of anisotropies at the bounce, leading to stronger isotropization of the universe. We now consider the ekpyrotic-like potential in the next subsection and compare its isotropization effects with the ekpyrotic potential presented in this section.

\subsection{Ekpyrotic-like potential }

\begin{figure}
\centering
\begin{subfigure}
    \centering
    \includegraphics[width=0.49\linewidth]{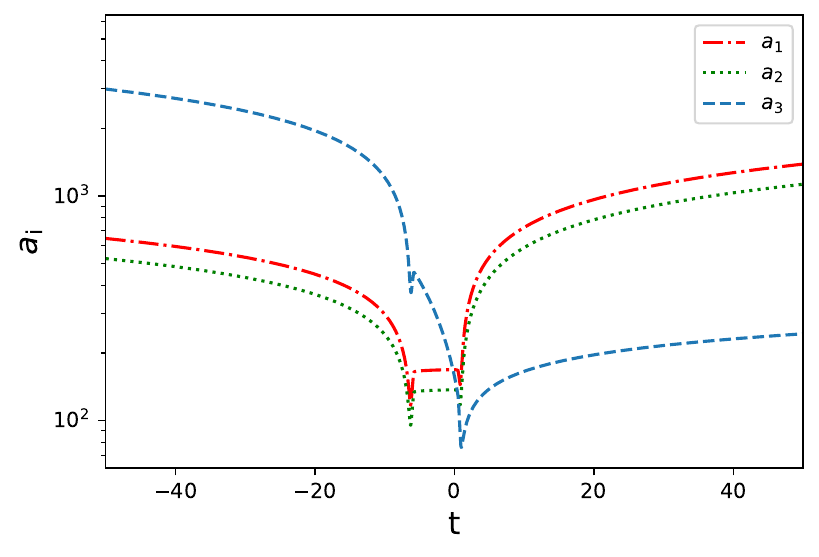}
\end{subfigure}
\begin{subfigure}
    \centering
    \includegraphics[width=0.49\linewidth]{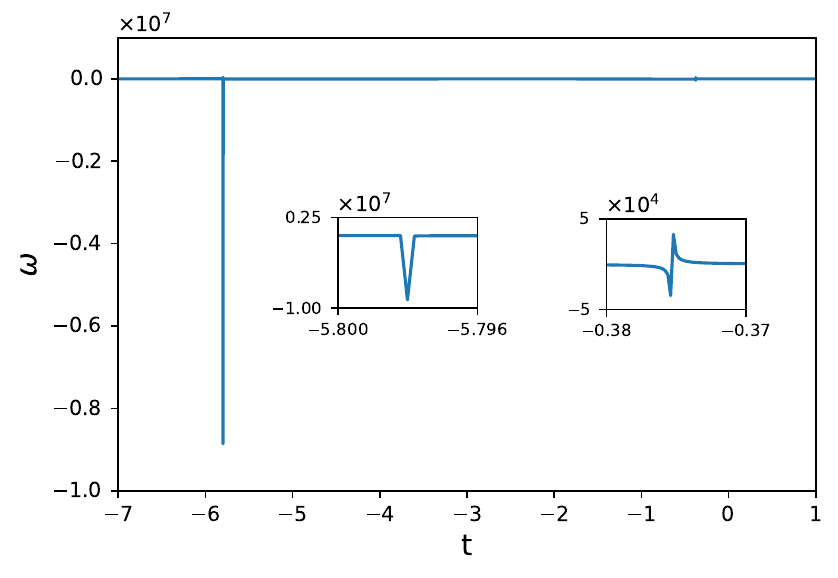}
\end{subfigure}
\begin{subfigure}
    \centering
    \includegraphics[width=0.49\linewidth]{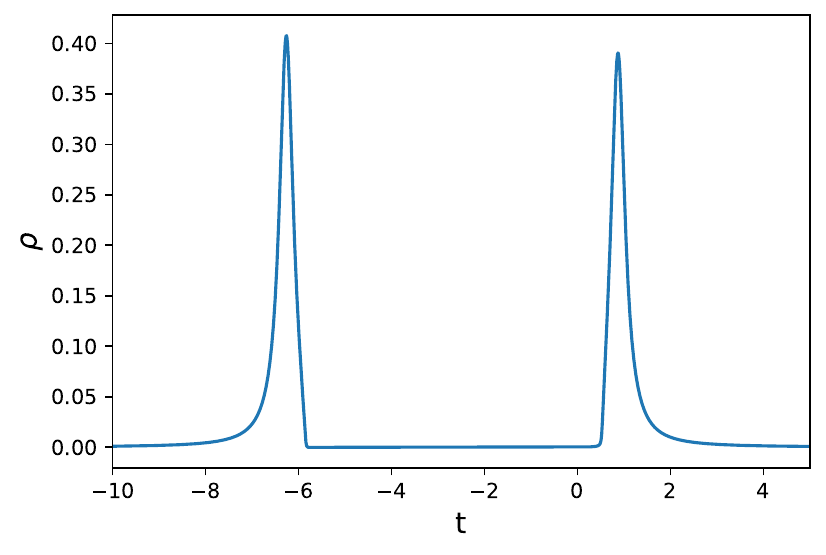}
\end{subfigure}
\begin{subfigure}
    \centering
    \includegraphics[width=0.49\linewidth]{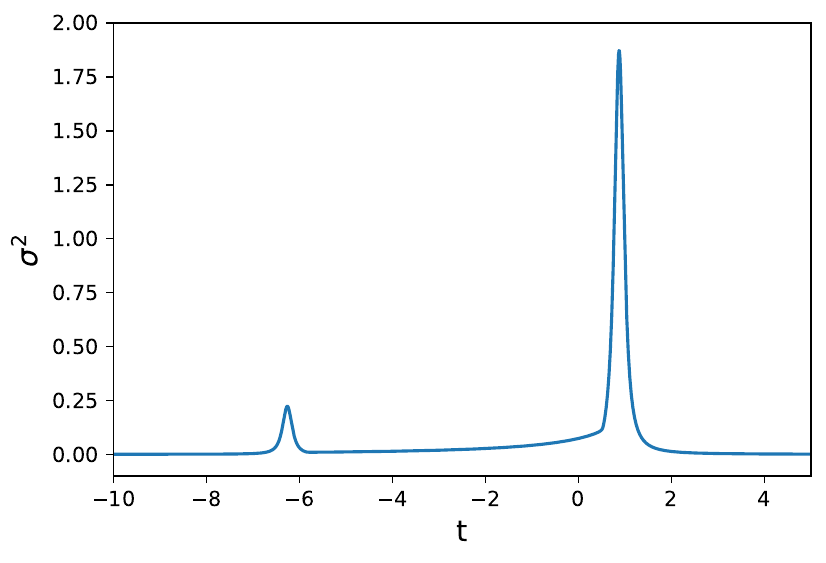}
\end{subfigure}
\caption{The typical evolution of directional scale factors, equation of state, energy density and anisotropic shear in the bounce regime for effective Bianchi-I spacetime coupled to the ekpyrotic-like scalar field with $U_0 = 0.25$. The number of bounces obtained is usually lower compared to the case of ekpyrotic field.}
\label{fig:Bianchi-I_EL_0.25_evolution}
\end{figure}

We now consider the effective Bianchi-I spacetime with an ekpyrotic-like potential, which is given by 
\begin{equation}\label{potential_EL}
U_{\mathrm{EL}}(\phi)=\frac{-2U_o}{e^{-\sqrt{\frac{16\pi}{p}}\phi}+e^{\beta\sqrt{\frac{16\pi}{p}}\phi}},
\end{equation}
with $U_o$, $p$ and $\beta$  all taking positive values. 
 For all simulations discussed, we have chosen the potential parameters $p=0.1$ and $\beta=5$. The typical evolution of the scale factors, energy density, shear scalar and the equation of state is shown in Fig. \ref{fig:Bianchi-I_EL_0.25_evolution} using initial conditions $c_1 = -0.11803, c_2 = -0.09536, p_1 = 21625, p_2 = 26592, p_3 = 23082, \phi=0.06151, \rho=7.87 \times 10^{-5}$ and $c_3$ solved to be $c_3 = -12.62885$ for $U_0 = 0.25$. Specifically, the singularity is resolved and the scale factors go through multiple bounces (although usually fewer than for the ekpyrotic potential) before entering an expanding phase leading to a large universe. As seen in the previous case, ekpyrosis ($w>1$) occurs only for short durations in the bounce regime. The energy density at the bounce is significantly increased due to the presence of an ekpyrotic phase in the bounce regime caused by the ekpyrotic-like potential, as described above. The shear scalar is also increased near the bounce, although, as before, it is not directly influenced by the potential so the effect is less easily predictable. The panel on the equation of state plot may appear to differ more from that of Fig. \ref{fig:Bianchi-I_Cycle_0.25_evolution}, as only one negative peak is visible. However, the positive peaks with $w>1$ responsible for ekpyrosis are still present, just of a smaller order of magnitude than this unusually large spike. This is verified by the second inset shown in the equation of state plot in \ref{fig:Bianchi-I_Cycle_0.25_evolution} as well as the presence of ekpyrosis in the results below.

Using the same methodology as in the previous subsection, we perform simulations of Bianchi-I universes with an ekpyrotic-like potential for a variety of initial conditions and compare the values of energy density and shear scalar at the last bounce with those obtained for the massless scalar field with matched initial conditions. In the first subsection we consider $U_0 = 0.25$, while in the next subsection we consider $U_0 = 1.25$ in order to see the effect of strengthening the ekpyrotic-like potential on isotropization. It should be noted that the strength of the ekpyrotic and ekpyrotic-like potentials have different strengths even when $U_0$ has the same magnitude, due to the other parameters and structure of the potential. 

\begin{figure}
\centering
\begin{subfigure}
       \centering
   \includegraphics[width=0.49\linewidth]
   {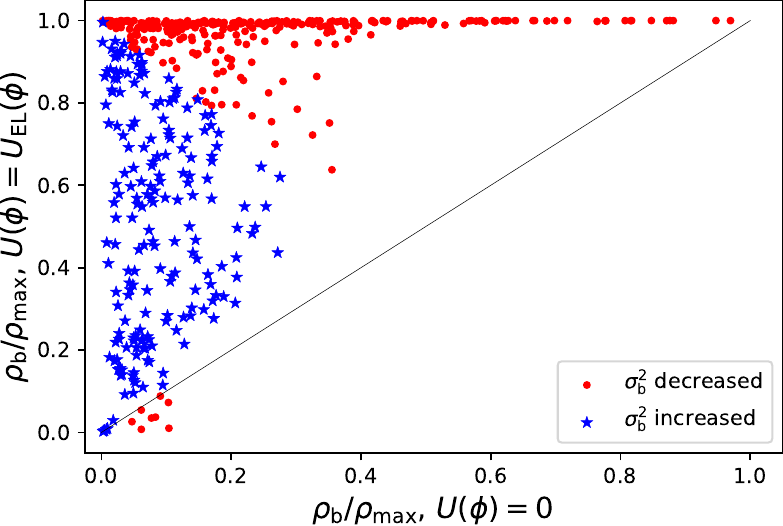}
\end{subfigure}
\begin{subfigure}
   \centering
   \includegraphics[width=0.49\linewidth]
   {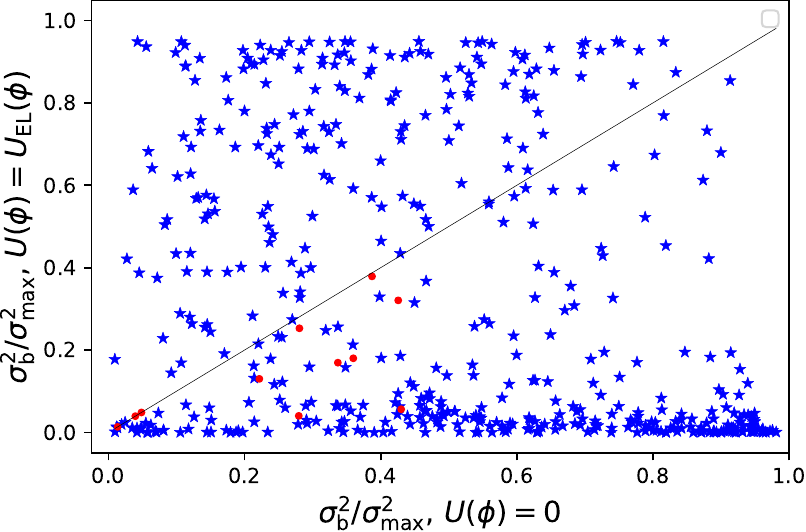}
\end{subfigure}
\begin{subfigure}
   \centering
   \includegraphics[width=0.49\linewidth]
   {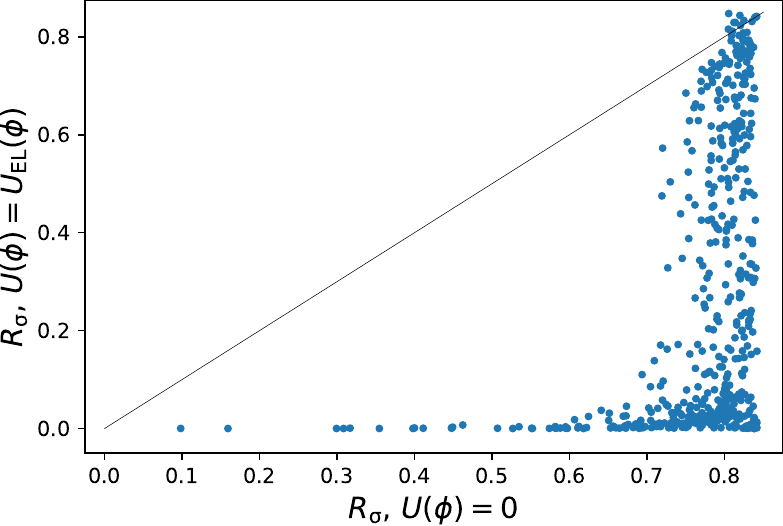}
\end{subfigure}
\caption{Bianchi-I effective spacetime with an ekpyrotic-like field ($U_0 = 0.25$): data from 500 simulations comparing energy density $\rho_\mathrm{b}/\rho_\mathrm{max}$, shear $\sigma_\mathrm{b}^2/\sigma_\mathrm{max}^2$, and the relative strength of anisotropies $R_\sigma$ at the last bounce between the ekpyrotic-like potential (y-axis) and the massless scalar case (x-axis) with matching initial conditions. The panels follow the same color coding format as described in Fig. \ref{fig:BianchiI_Cycle_0.25}. In contrast to the ekpyrotic case, not all simulations decrease with respect to the massless scalar case in the $R_\sigma$ plot.}

\label{fig:BianchiI_EL_Cycle_0.25}
\end{figure}

\subsubsection{Effective Bianchi-I spacetime with ekpyrotic-like potential: $U_0=0.25$}

Fig. \ref{fig:BianchiI_EL_Cycle_0.25} shows the comparison of the energy density and shear scalar at the last bounce with their respective values in the case of a massless scalar field for $U_0 = 0.25$. We find that as in the case of the ekpyrotic potential considered above, the ekpyrotic-like potential also leads to isotropization at the bounce. \textcolor{black}{Note that in comparison to the simulation for the ekpyrotic potential shown in Fig. 1, the bounce appears smoother. Though, there is no general reason for the ekpyrotic-like potential to give a smoother bounce, the contrast between Fig. 1 and Fig. \ref{fig:Bianchi-I_EL_0.25_evolution} is due to the difference in initial conditions, especially  of the potential and the resulting impact on the equation of state. In the ekpyrotic-like case shown here the potential is about 100 times larger in magnitude than the case shown in Fig. 1 for a similar value of initial energy density. Further, the the initial equation of state at $t=0$ is highly stiff ($w \approx 13.86$) compared to the ekpyrotic potential case in Fig. 1 where it is $w \approx 0.82$. We suspect that the initial equation of state being highly stiff causes  the bounce to be more isotropic and smoother than the case corresponding to Fig. 1.}

In the ekpyrotic-like case, the energy density at the bounce is increased compared to the massless case in all but a few simulations. However, this is not necessarily accompanied by a decrease in the shear scalar, as seen by the color coding in the energy density plot in Fig. \ref{fig:BianchiI_EL_Cycle_0.25} where only the red points experienced $\sigma_\mathrm{b}^2$ decrease. The second panel on the comparison of the shear scalar confirms the same, although $61.00\%$ of the simulations did have $\sigma_\mathrm{b}^2$ decrease compared to a massless scalar. Yet, the relative strength of the anisotropies, represented by the ratio $R_\sigma$ at the bounce, is decreased by the ekpyrotic-like potential in all but a few cases, as seen in the third panel of Fig. \ref{fig:BianchiI_EL_Cycle_0.25}. Moreover, similar to the case of effective Bianchi-I spacetime with ekpyrotic potential, there is no mixing of the red and blue dots in the energy density plots indicating a clear boundary between the two populations of simulations.

There are a few differences with the ekpyrotic case which should be noted. In contrast to the ekpyrotic potential, the isotropization effect of the ekpyrotic-like potential seems to be milder. Unlike the ekpyrotic case with low strength of the potential, the dots are more spread out and there is no clustering of the dots into two groups in all three plots in Fig. \ref{fig:BianchiI_EL_Cycle_0.25} in contrast to Fig. \ref{fig:BianchiI_Cycle_0.25}. Moreover, there exist a very small number of outliers in this case, i.e. simulations where isotropization is not achieved at all. However, these occur only when the matching massless scalar simulation was already very isotropic, and it remains the case that the ekpyrotic-like potential leads to isotropization of the universe at the bounce in $97.40\%$ of the simulations.

\subsubsection{Effective Bianchi-I spacetime with ekpyrotic-like potential: $U_0=1.25$}

\begin{figure}
\centering
\begin{subfigure}
       \centering
   \includegraphics[width=0.49\linewidth]
   {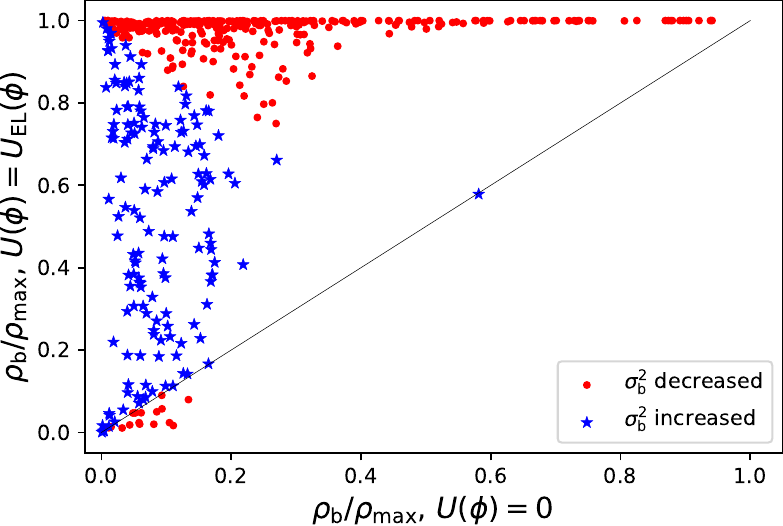}
\end{subfigure}
\begin{subfigure}
   \centering
   \includegraphics[width=0.49\linewidth]
   {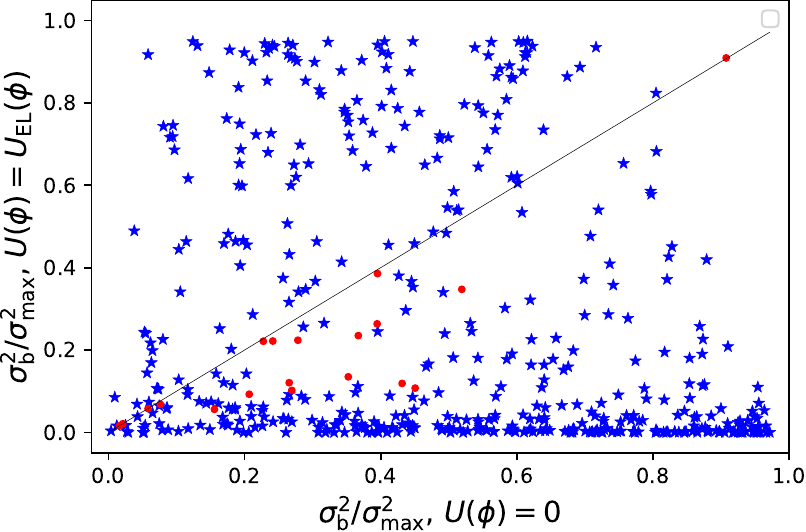}
\end{subfigure}
\begin{subfigure}
   \centering
   \includegraphics[width=0.49\linewidth]
   {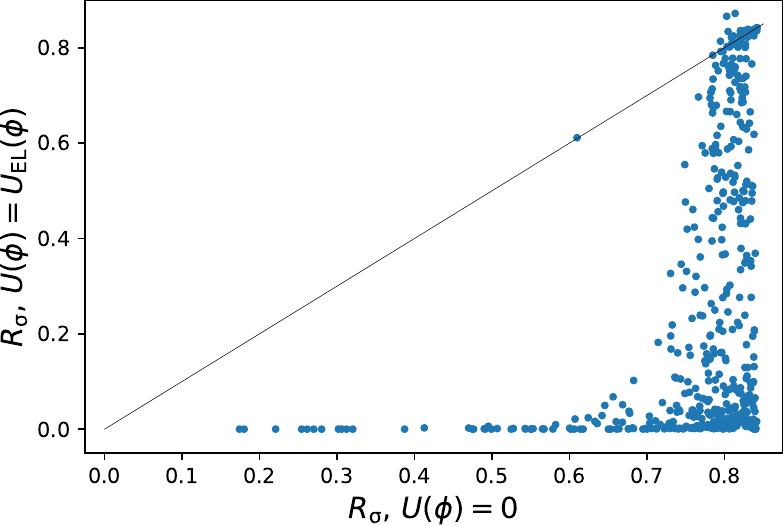}
\end{subfigure}

\caption{Bianchi-I effective spacetime with an ekpyrotic-like field ($U_0 = 1.25$): comparing energy density $\rho_\mathrm{b}/\rho_\mathrm{max}$, shear $\sigma_\mathrm{b}^2/\sigma_\mathrm{max}^2$, and the relative strength of anisotropies $R_\sigma$ at the last bounce between the ekpyrotic-like potential (y-axis) and the massless scalar case (x-axis) with matching initial conditions for 500 simulations. The color coding scheme described in Fig. \ref{fig:BianchiI_Cycle_0.25} is used to provide additional information. Varying the strength of the ekpyrotic-like potential does not yield drastic changes.}

\label{fig:BianchiI_EL_1.25}
\end{figure}

We now study the effect of increasing the strength of the ekpyrotic potential on the isotropization achieved by it at the bounce. We study this in Fig. \ref{fig:BianchiI_EL_1.25}, which shows the color-coded comparison plots for the ekpyrotic-like potential with $U_0=1.25$. Based on our intuition from the previous subsection on ekpyrotic potential, where a strong increase in isotropization was seen with the increase in the strength of the potential, one would expect similar results in the case of ekpyrotic-like potential. However, as can be seen by comparing the plots in Fig. \ref{fig:BianchiI_EL_Cycle_0.25} with the corresponding ones in Fig. \ref{fig:BianchiI_EL_1.25}, there is no noticeable increase in the isotropization achieved by increasing the strength of the ekpyrotic-like potential. To further investigate, we increased the strength of the ekpyrotic-like potential to $U_0=12.5$ and $U_0=125$ and studied the cases by performing $500$ simulations each, as done for all the cases in this manuscript. Still, no substantial increase in isotropization achieved by the ekpyrotic-like potential was observed.

The extensive numerical simulations in this section on the effective Bianchi-I spacetime establish that both the ekpyrotic and ekpyrotic-like scalar fields lead to a dilution of the anisotropies in the bounce regime, causing a degree of isotropization of the universe dependent on the strength of the potential. However, even qualitatively it can be seen that the effects are much stronger in the case of an ekpyrotic field in comparison to the ekpyrotic-like potential. Further, the isotropization caused by the ekpyrotic field responds with much more sensitivity to an increase in the strength of the potential compared to the ekpyrotic-like field. Further, we find that the universe undergoes several short-lived bounces in a rapid succession, which are accompanied by short spurts of ekpyrosis, which are ultimately responsible for the isotropization caused. Importantly, we see that ekpyrosis only directly affects the energy density, rapidly pumping it up in the approach to the bounce, which is the primary cause of the isotropization. Its effect on anisotropies is indirect and not as consistent. The changes caused in anisotropies are at best a secondary contributory factor to isotropization. We now explore the effective Bianchi-IX dynamics to study the isotropization effects of these ekpyrotic and ekpyrotic-like potentials.

\section{Ekpyrosis in Bianchi-IX spacetime in LQC}

In this section, we will consider Bianchi-IX spacetimes with ekpyrotic and ekpyrotic-like potentials. In particular, we simulate matching pairs of universes with ekpyrotic or ekpyrotic-like potentials and universes with only a massless scalar field. The initial conditions of the pairs of simulations are the same (up to solving for $c_3$), so we can compare the evolution of the energy density and anisotropic shear with and without the potential present using the same methods as in the Bianchi-I section. In particular, we wish to determine whether ekpyrosis leads to isotropization. In the first subsection, we will consider Bianchi-IX spacetimes coupled to ekpyrotic scalar fields, and in the second Bianchi-IX spacetimes coupled to ekpyrotic-like scalar fields. We consider two different strengths of both potentials. Finally, we discuss how the results compare with those of the prior section, noting that the effective dynamics of Bianchi-I spacetimes are a limiting case of the effective dynamics of Bianchi-IX spacetimes in `K' quantization. Throughout, the potential and simulation parameters are chosen to be the same as for the Bianchi-I simulations.

\subsection{Ekpyrotic potential}

\begin{figure}
\centering
\begin{subfigure}
    \centering
    \includegraphics[width=0.49\linewidth]
    {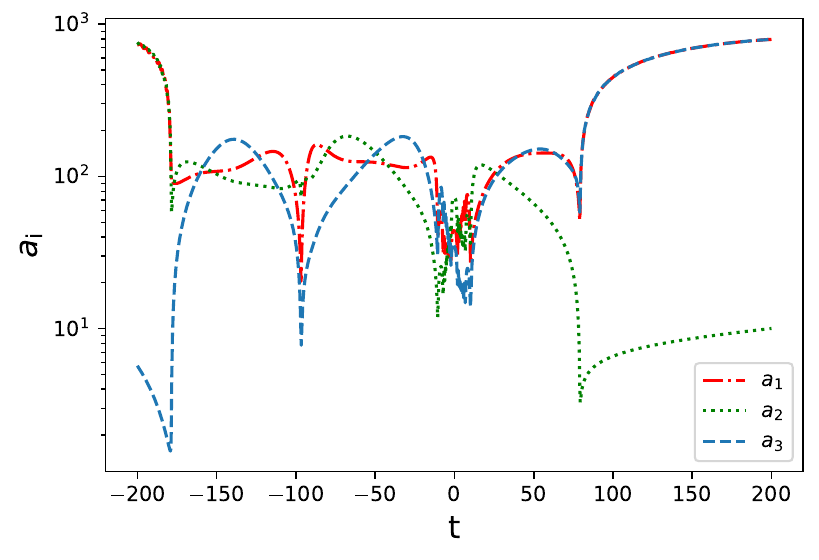}
\end{subfigure}
\begin{subfigure}
    \centering
    \includegraphics[width=0.49\linewidth]
    {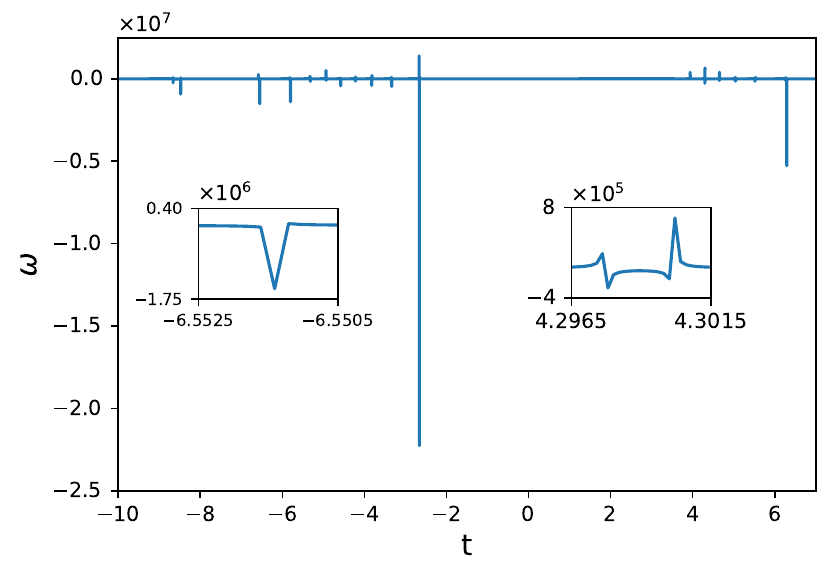}
\end{subfigure}
\begin{subfigure}
    \centering
    \includegraphics[width=0.49\linewidth]
    {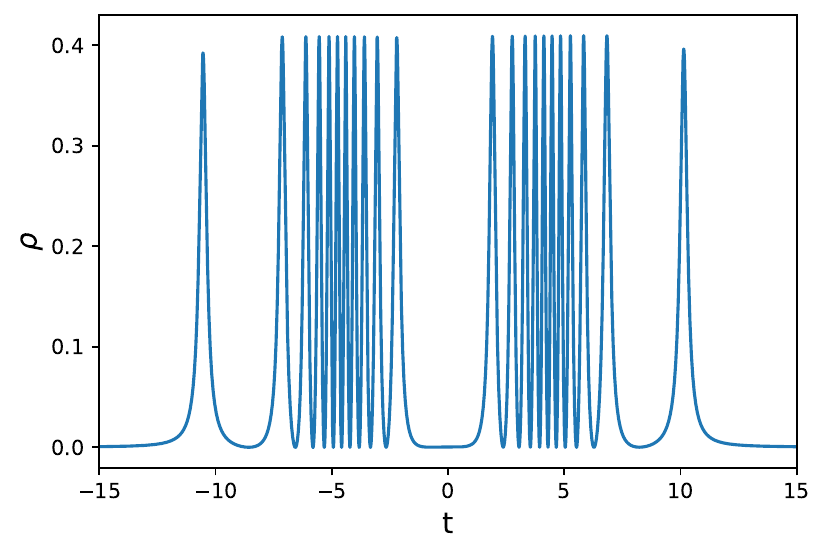}
\end{subfigure}
\begin{subfigure}
    \centering
    \includegraphics[width=0.49\linewidth]
    {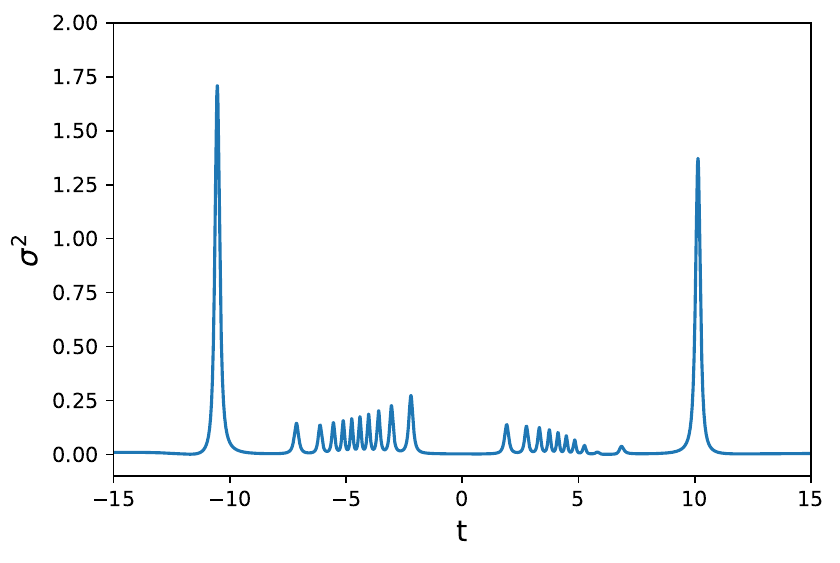}
\end{subfigure}
\caption{An example simulation showing the evolution of directional scale factors, equation of state, energy density, and the anisotropic shear for effective Bianchi-IX spacetime coupled to the ekpyrotic scalar field with $U_0 = 0.25$. As seen here, the universe goes through multiple bounces before entering a macroscopic semi-classical expanding phase.}
\label{fig:BianchiIX_Cycle_0.25_evolv}
\end{figure}

We will first consider Bianchi-IX spacetimes with an ekpyrotic potential given by equation \eqref{Ekpyrotic_potential}. Typical evolution of the scale factors, energy density, shear scalar, and the equation of state  is shown in Figure \ref{fig:BianchiIX_Cycle_0.25_evolv} using initial conditions $c_1 = -0.16267, c_2 = -0.19149, p_1 = 18324, p_2 = 11144, p_3 = 23155, \phi=5.54 \times 10^{-6}, \rho=9.58 \times 10^{-5}$ and $c_3$ solved to be $c_3 = -1.26138$ at $U_0 = 0.25$. As in the effective dynamics of loop quantized Bianchi-I spacetimes, we see that the singularity is resolved and the scale factors go through numerous bounces. In fact, the evolution of the scale factors is significantly more complex than the typical behavior in Bianchi-I spacetimes, reflecting the greater complexity of Bianchi-IX dynamics. This also leads to difficulty in choosing a time scale for simulations that leads to a large universe phase after the last bounce.  Similar to the Bianchi-I spacetime case, the energy density and shear scalar both peak around the bounce locations, with a clearer direct effect of bounces on the energy density. Finally, the equation of state plot appears largely similar to that of the Bianchi-I case with an ekpyrotic potential. In particular, we see that the positive peaks with $w>1$ required for ekpyrosis are present. Analogous to the Bianchi-I case, ekpyrosis occurs in multiple short durations over multiple bounces in the bounce regime.

Using the same methodology as in the previous section for Bianchi-I spacetimes, we perform simulations of Bianchi-IX universes with an ekpyrotic potential for a variety of initial conditions. We then compare the values of energy density and the anisotropic shear at the last bounce with those obtained for the massless scalar field with matched initial conditions. In the first subsection we consider $U_0=0.25$ and in the second subsection $U_0=1.25$, to compare the effect of strengthening the potential.

\begin{figure}
\centering
\begin{subfigure}
       \centering
   \includegraphics[width=0.49\linewidth]{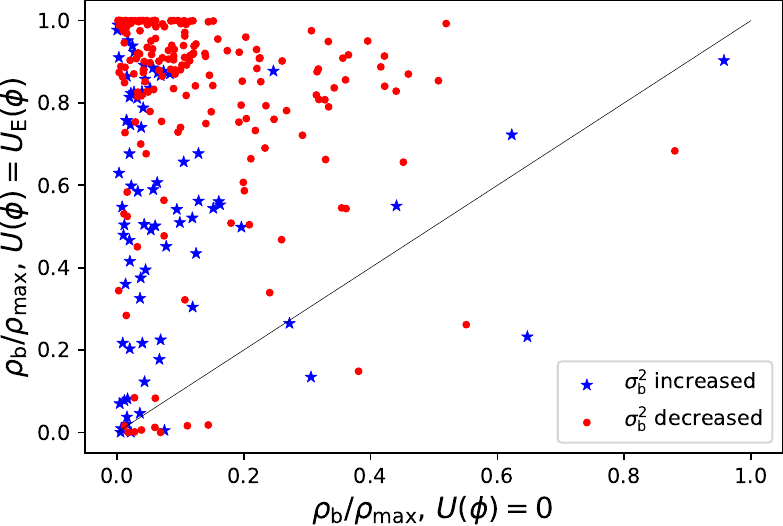}
\end{subfigure}
\begin{subfigure}
   \centering
   \includegraphics[width=0.49\linewidth]{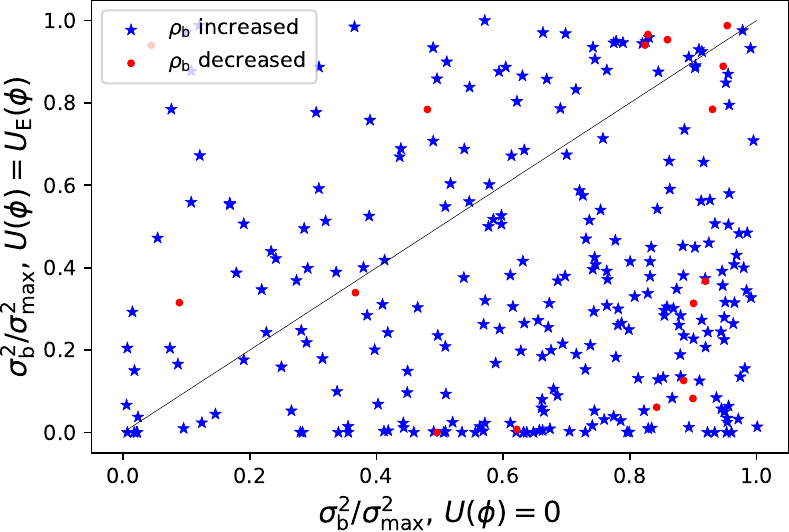}
\end{subfigure}

\begin{subfigure}
    \centering
    \includegraphics[width=0.49\linewidth]{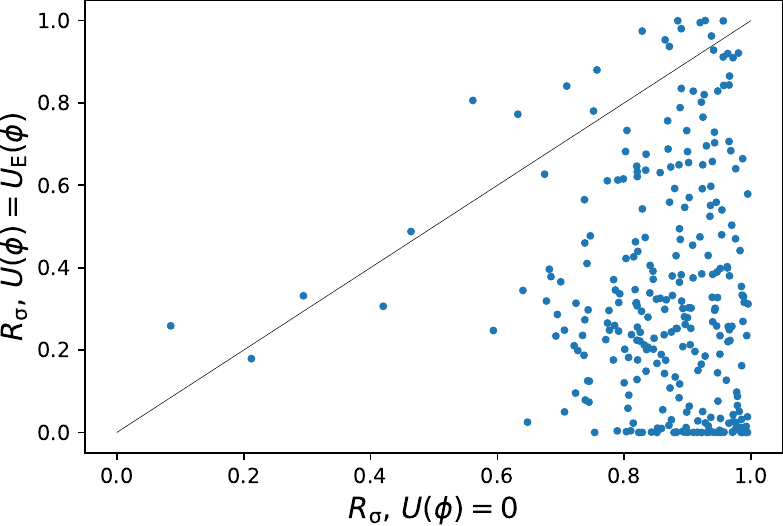}
\end{subfigure}
\caption{Bianchi-IX effective spacetime with an ekpyrotic field ($U_0 = 0.25$): comparison of energy density $\rho_\mathrm{b}/\rho_\mathrm{max}$, shear $\sigma_\mathrm{b}^2/\sigma_\mathrm{max}^2$, and the relative strength of anisotropies $R_\sigma$ at the last bounce between the ekpyrotic potential (y-axis) and the massless scalar case (x-axis) with matching initial conditions for 500 simulations. The color coding scheme used in previous comparison plots is used. In contrast to the
Bianchi-I case, not all simulations decrease with respect to the massless scalar case in the $R_\sigma$ plot.}
\label{fig:BianchiIX_Cycle_0.25}
\end{figure}

\subsubsection{Effective Bianchi-IX spacetime with ekpyrotic potential: $U_0=0.25$}

For $U_0=0.25$, Fig. \ref{fig:BianchiIX_Cycle_0.25} compares the values of energy density and shear scalar at the last bounce in the time range of the simulations with their respective values for a massless scalar field. As before, the points are color coded to provide additional information about the simulations. The first panel shows $\rho_\mathrm{b}/\rho_\mathrm{max}$ for the ekpyrotic potential compared to the massless scalar field with matching initial conditions, and we note that almost all the simulations show an increase in the energy density. However, there are several key differences from Fig. \ref{fig:BianchiI_Cycle_0.25} for the Bianchi-I case with the same potential. First, while every simulation had the energy density increase under the ekpyrotic potential in Bianchi-I spacetimes, a number of simulations did not have the energy density increase in Bianchi-IX spacetimes. Additionally, the clear boundary of separation between points where the shear scalar increased (blue) or decreased (red) is no longer present. In general, we see that the change in energy density is more spread, whether or not it is considered in relation to the change in anisotropic shear. In fact, no clear clustering of points occurs. While the effect of ekpyrosis is more varied for the Bianchi-IX potential and the average energy density of the massless scalar field in Bianchi-IX spacetimes is lower than in Bianchi-I spacetimes, the average value of $\rho_\mathrm{b}/\rho_\mathrm{max}$ is $\rho_\mathrm{av} = 0.303$, which is larger than the Bianchi-I spacetime under the same ekpyrotic potential. Hence, it seems the ekpyrotic potential may have a stronger, if more varied, effect on energy density in Bianchi-IX spacetimes than in Bianchi-I.

The second panel of Fig. \ref{fig:BianchiIX_Cycle_0.25} shows the value of the anisotropic shear scalar $\sigma_\mathrm{b}^2/\sigma_\mathrm{max}^2$ at the last bounce compared to the massless scalar field at the bounce with matching initial conditions. While the anisotropic shear is not always decreased, a much higher proportion ($71.99\%$) of simulations found a decrease in the anisotropic shear than in Bianchi-I spacetimes where only $37.6\%$ of simulations had the anisotropic shear decrease. Similar to the behavior of energy density, no clustering of points is seen unlike in the corresponding panel of Fig. \ref{fig:BianchiI_Cycle_0.25}. However, we again notice that an increase in energy density at the bounce is not necessarily correlated with a decrease in the anisotropic shear, as almost every simulation had the energy density increase under the potential and many fewer had the anisotropic shear decrease. 
 Fig. \ref{fig:BianchiIX_Cycle_0.25} indicates that the addition of the ekpyrotic potential tends to decrease the average anisotropic shear, in contrast to the Bianchi-I case where this effect was absent. However, before concluding that there is a stronger isotropization effect in Bianchi-IX spacetimes, we should consider the differing responses of the energy density and anisotropic shear.

The third panel of Fig. \ref{fig:BianchiIX_Cycle_0.25} compares the value of the ratio $R_\sigma$ at the last bounce for the ekpyrotic potential and massless scalar field. While most points in these figures are below the diagonal line ($93.97\%$ to be exact), this differs from the Bianchi-I universe with an ekpyrotic potential where every single simulation conducted resulted in a decrease of this ratio. However, we see a higher density of simulations at the bottom of the figure as opposed to the cluster of points near the diagonal line in the Bianchi-I case. This may indicate a stronger, if less consistent, effect of ekpyrosis on isotropization in Bianchi-IX universes. In fact, similar to the results for energy density, the average value of $R_\sigma$ is smaller under the ekpyrotic potential in Bianchi-IX simulations than in Bianchi-I despite a larger value in Bianchi-IX for a massless scalar field. Hence, we observe that isotropization is almost always achieved in the presence of an ekpyrotic field in comparison to the massless scalar field.

Finally, before moving to the case with an increased potential strength, we note that the relationship between energy density and anisotropic shear is not yet well understood for loop quantized Bianchi-IX spacetimes. We discussed above an inverse parabolic relationship between anisotropic shear and energy that was found numerically for Bianchi-I spacetimes. However, as of yet no such relationship has been established for Bianchi-IX spacetimes in LQC. Hence, we are unable to provide a similar explanation for the anisotropic shear failing to decrease in instances where the energy density increases. We conclude by noting that, as we have taken the `K' quantization of Bianchi-IX spacetime the Bianchi-I effective field theory is a limiting case of the Bianchi-I effective field theory. Hence, we expect a more complex relationship between the anisotropic shear and energy density for Bianchi-IX universes, which contains the established relationship for Bianchi-I universes.

\subsubsection{Effective Bianchi-IX spacetime with ekpyrotic potential: $U_0=1.25$}

\begin{figure}
\centering
\begin{subfigure}
       \centering
   \includegraphics[width=0.49\linewidth]
   {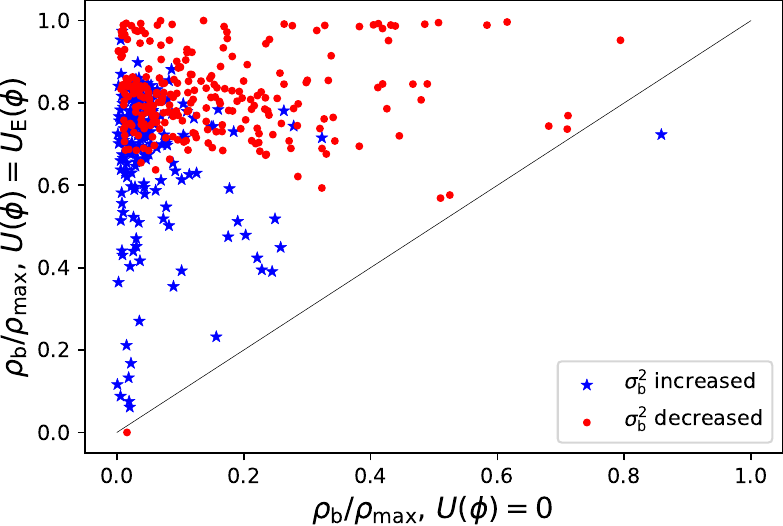}
   \label{fig:BianchiIX_Cycle_1.25_rho_compare_shear_marked}
\end{subfigure}
\begin{subfigure}
   \centering
   \includegraphics[width=0.49\linewidth]
   {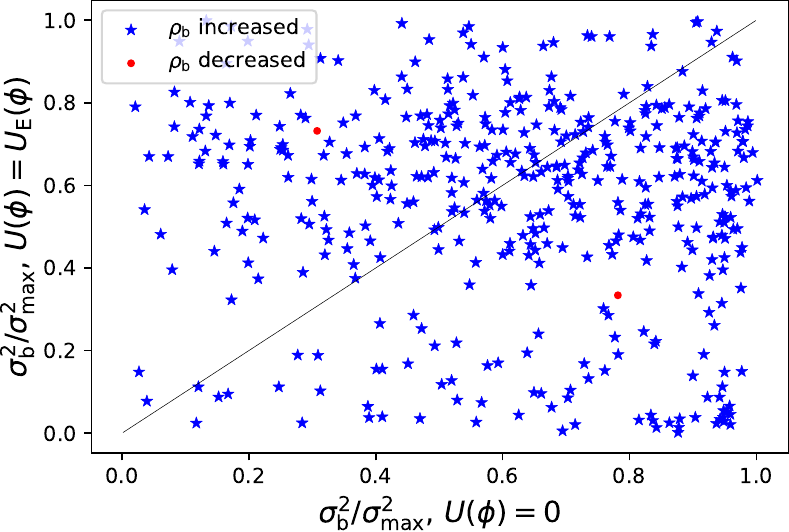}

   \label{fig:BianchiIX_Cycle_1.25_shear_compare_rho_marked}
\end{subfigure}

\begin{subfigure}
   \centering
   \includegraphics[width=0.49\linewidth]
   {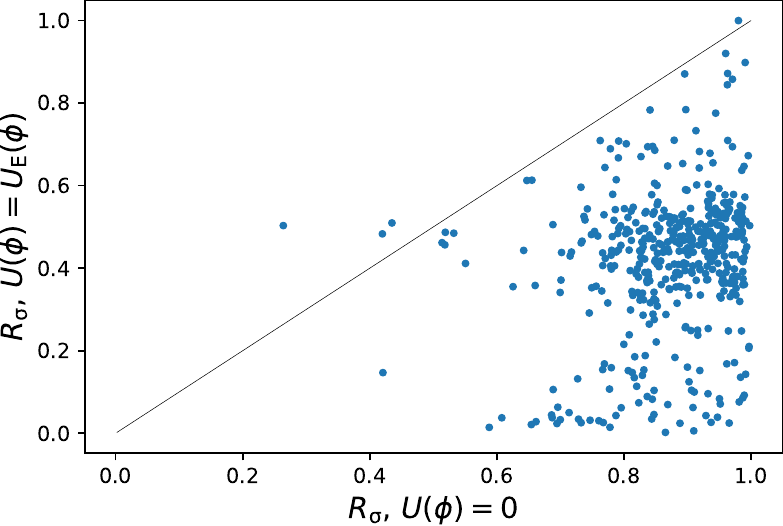}

   \label{fig:BianchiIX_Cycle_1.25_shear_by_rho&shear_compare}
\end{subfigure}

\caption{Bianchi-IX effective spacetime with an ekpyrotic field ($U_0 = 1.25$): energy density $\rho_\mathrm{b}/\rho_\mathrm{max}$, shear $\sigma_\mathrm{b}^2/\sigma_\mathrm{max}^2$, and the relative strength of anisotropies $R_\sigma$ at the last bounce for the ekpyrotic potential (y-axis) compared with the massless scalar case (x-axis) with matching initial conditions for 500 simulations. The panels follow the same format as described in Fig. \ref{fig:BianchiI_Cycle_0.25}. When comparing to the $U_0 = 0.25$ case in Fig. \ref{fig:BianchiIX_Cycle_0.25}, it is clear that increasing the strength of the ekpyrotic field increases the number of simulations where $\rho_\mathrm{b}/\rho_\mathrm{max}$ increases with respect to the massless scalar field.} 

\label{fig:BianchiIX_Cycle_1.25}
\end{figure}

Having studied the ekpyrotic potential and its effects on isotropization of Bianchi-IX universes, we now consider the effect of strengthening the potential. In particular, we increase the potential parameter $U_0$ from $U_0=0.25$ to $U_0=1.25$. In Fig. \ref{fig:BianchiIX_Cycle_1.25} we plot the effects of the ekpyrotic potential with $U_0=1.25$ by comparing the value of tbe energy density, anisotropic shear, and $R_\sigma$ with the massless scalar field for matched initial conditions as described above. The arguments for Fig. \ref{fig:BianchiIX_Cycle_0.25} regarding isotropization and the behavior of the energy density and anisotropic shear at the last bounce mostly generalize to this stronger potential case. However, we find that there are several surprising differences. In the first panel of Fig. \ref{fig:BianchiIX_Cycle_1.25} we see that the energy density has increased under the ekpyrotic potential in all but a few exceptions. Further, the average value of the energy density under the ekpyrotic potential with $U_0=1.25$ is found to be greater than that observed in the $U_0=0.25$ case. This effect of increasing the ekpyrotic potential on the energy density is not surprising. However, we also note a trend towards regaining a partial separation of the simulations into regions where the anisotropic shear increased (red) or decreased (blue) that was seen for Bianchi-I simulations using effective dynamics of LQC. While there is no clear separation of the points unlike in the Bianchi-I case, this may still hint towards an interesting relationship between anisotropic shear and energy density in Bianchi-IX universes.

The anisotropic shear is studied in the second panel of Fig. \ref{fig:BianchiIX_Cycle_1.25}. In contrast to what one might expect, raising the ekpyrotic potential to $U_0=1.25$ led to the number of simulations above the diagonal increasing to $56.26\%$. Further, the average value of the anisotropic shear at the last bounce was found to be $\sigma_\mathrm{avg} = 6.53$, as opposed to $4.30$ in the $U_0=0.25$ case. In fact, this effect is visible due to the large number of simulations gathered near the center right of the figure, rather than near the bottom of the figure as in Fig. \ref{fig:BianchiIX_Cycle_0.25}. This is counter to the expectation that a stronger ekpyrotic potential would lead to greater isotropization of the shear scalar. However, as changes in energy density and anisotropic shear have opposite impacts on the total isotropization, we should again consider the ratio $R_\sigma$. This is done in the third panel of Fig. \ref{fig:BianchiIX_Cycle_1.25}, where we confirm a strong isotropization effect by finding that $99.12\%$ of the simulations fall below the diagonal line. However, the points no longer cluster at the bottom of the figure, resulting in the average value of $R_\sigma$ increasing from $0.34$ for $U_0=0.25$ to $0.41$ for $U_0=1.25$. This effect of increasing the potential strength leading to a larger average value of $R_\sigma$ is surprising it is opposite to the one seen for Bianchi-I universes.

\subsection{Ekpyrotic-like potential }

\begin{figure}
\centering
\begin{subfigure}
    \centering
\includegraphics[width=0.47\linewidth]
    {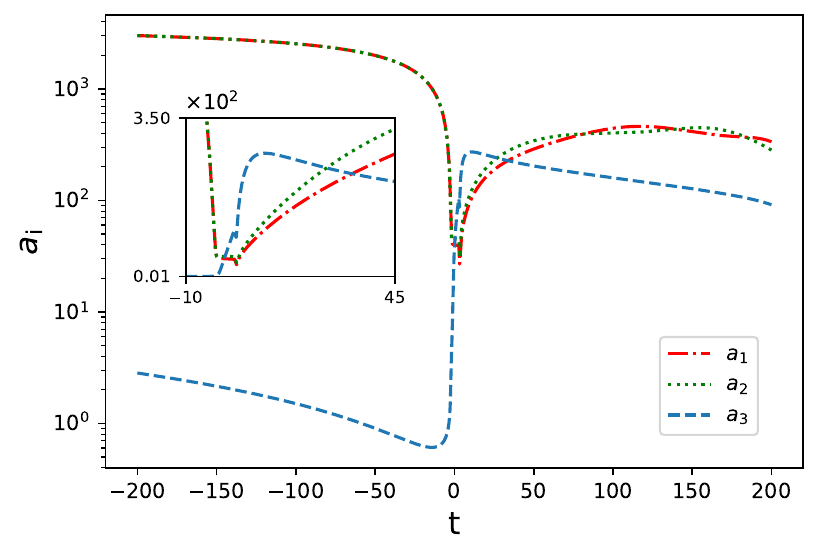}
 \end{subfigure}
\begin{subfigure}
    \centering
    \includegraphics[width=0.49\linewidth]
    {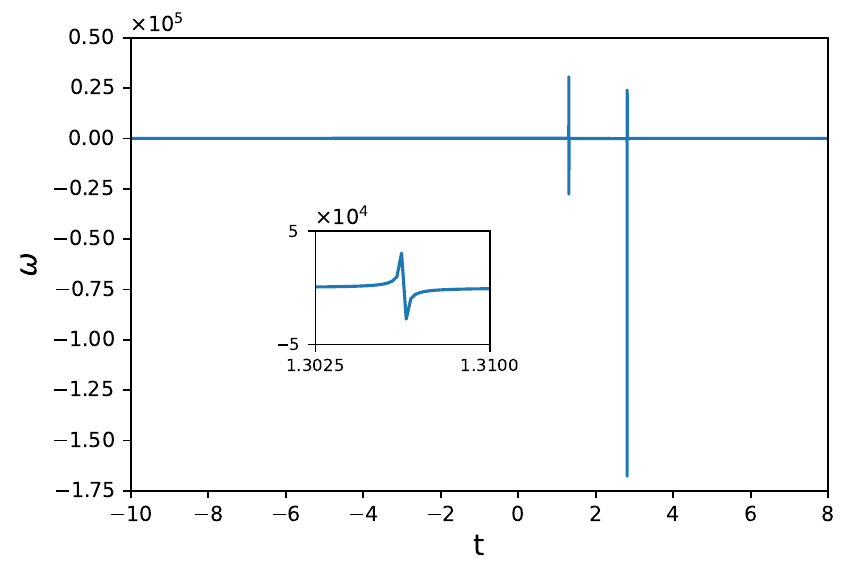}
\end{subfigure}
\begin{subfigure}
    \centering
    \includegraphics[width=0.49\linewidth]
    {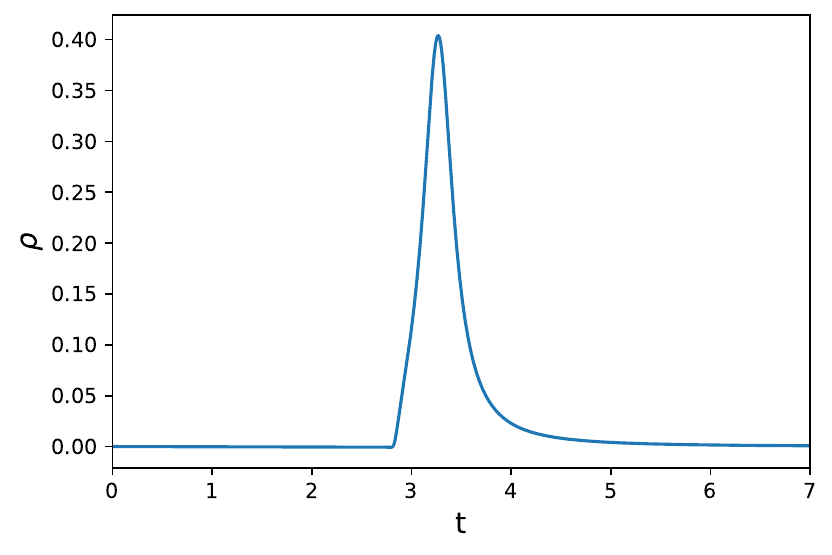}
\end{subfigure}
\begin{subfigure}
    \centering
    \includegraphics[width=0.47\linewidth]
    {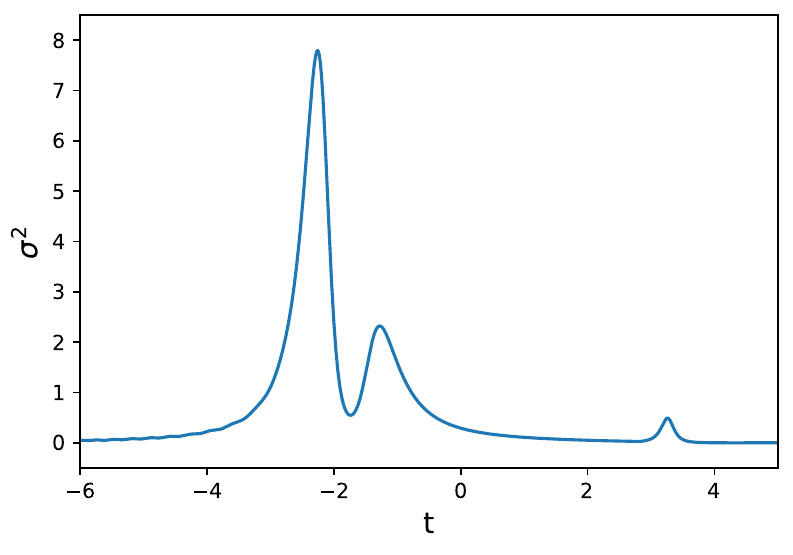}
\end{subfigure}
\caption{The typical evolution of directional scale factors, equation of state, energy density, and the anisotropic shear in the bounce regime for effective Bianchi-IX spacetime coupled to the ekpyrotic-like scalar field with $U_0 = 0.25$. As noticed earlier for the Bianchi-I spacetime, the number of bounces with ekpyrotic-like potential is generally lower in comparison with ekpyrotic field in case of Bianchi-IX effective spacetime as well.} 

\label{fig:Bianchi-IX_EL_0.25_evol}
\end{figure}

We now consider Bianchi-IX spacetimes with an ekpyrotic-like potential given by equation \eqref{potential_EL}. The typical evolution of the scale factors, energy density, anisotropic shear scalar, and equation of state are shown in Fig. \ref{fig:Bianchi-IX_EL_0.25_evol} using initial condition $c_1 = -0.20028, c_2 = -0.01286, p_1 = 12274, p_2 = 10855, p_3 = 13173, \phi=0.09425, \rho=9.56 \times 10^{-5}$ and $c_3$ solved to be $c_3 = 16.03946$ at $U_0 = 0.25$. As in all cases above, the singularity is resolved and the scale factors undergo several quantum gravitation bounces before entering a large universe regime. However, the scale factors undergo fewer bounces and the behavior is far less chaotic than under the ekpyrotic potential, similarly to the effect found in Bianchi-I universes. Again, the energy density and anisotropic shear peak around the bounces, and in this particular simulation there is clearly some isotropization between the first and last bounces. The peaks in the energy density near the bounce are directly a consequence of the equation of state becoming ultra-stiff at the bounce, where the peaks satisfy $w>1$. Consistent with all the previous cases, we find again that ekpyrosis occurs in short spurts spread over multiple bounces in the bounce regime.

To understand the effect of the ekpyrotic-like potential on isotropization through comparison with the massless scalar field case we consider potentials with both $U_0=0.25$ and $U_0=1.25$ in order to see the effect of strengthening the ekpyrotic-like potential on isotropization. This is explored in Figures~\ref{fig:BianchiIX_EL_0.25}~and~\ref{fig:BianchiIX_EL_1.25}, where the values of $\rho_\mathrm{b}/\rho_\mathrm{max}$, $\sigma_\mathrm{b}^2/\sigma_\mathrm{max}^2$ and $R_\sigma$ are plotted for simulations with and without the ekpyrotic-like potential with matching initial conditions (up to solving for $c_3$). It is again important to note that setting the same value for $U_0$ does not correspond to setting the strength of an ekpyrotic and ekpyrotic-like potential to be equal. Hence, we may only qualitatively compare the results for ekpyrotic and ekpyrotic-like potentials. However, we may certainly compare the effects of the same potential in Bianchi-I and Bianchi-IX spacetimes.

\subsubsection{Effective Bianchi-IX spacetime  with ekpyrotic-like potential: $U_0=0.25$}

\begin{figure}
\centering
\begin{subfigure}
       \centering
   \includegraphics[width=0.49\linewidth]
   {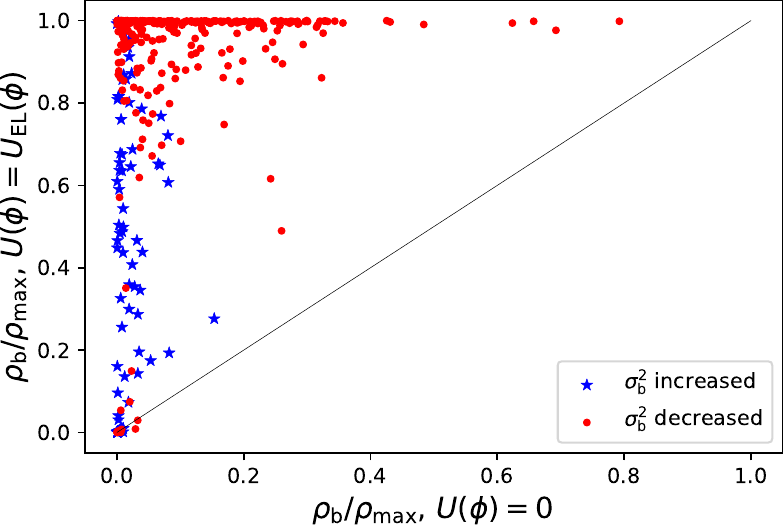}
   \label{fig:BianchiIX_EL_0.25_rho_compare_shear_marked}
\end{subfigure}
\begin{subfigure}
   \centering
   \includegraphics[width=0.49\linewidth]
   {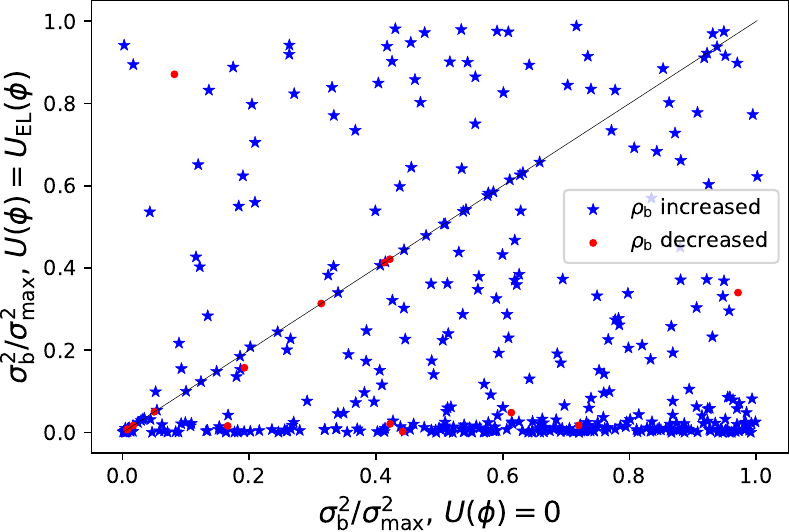}

   \label{fig:BianchiIX_EL_0.25_shear_compare_rho_marked}
\end{subfigure}

\begin{subfigure}
   \centering
   \includegraphics[width=0.49\linewidth]
   {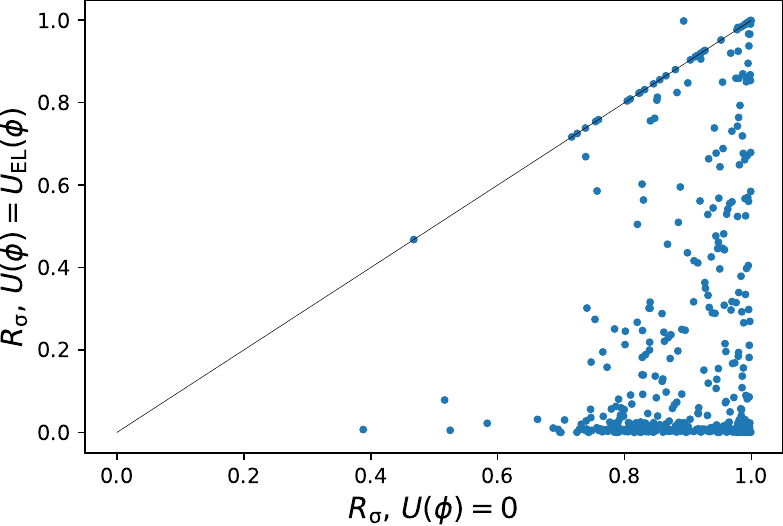}

\end{subfigure}

\caption{Bianchi-IX effective spacetime with an ekpyrotic-like field ($U_0 = 0.25$): 500 simulations of energy density $\rho_\mathrm{b}/\rho_\mathrm{max}$, shear $\sigma_\mathrm{b}^2/\sigma_\mathrm{max}^2$, and the relative strength of anisotropies $R_\sigma$ at the last bounce for the ekpyrotic-like potential (y-axis) compared with the massless scalar case (x-axis) with matching initial conditions. The panels follow the same format as described in Fig. \ref{fig:BianchiI_Cycle_0.25}.} 

\label{fig:BianchiIX_EL_0.25}
\end{figure}

\begin{figure}
\centering
\begin{subfigure}
       \centering
   \includegraphics[width=0.49\linewidth]
   {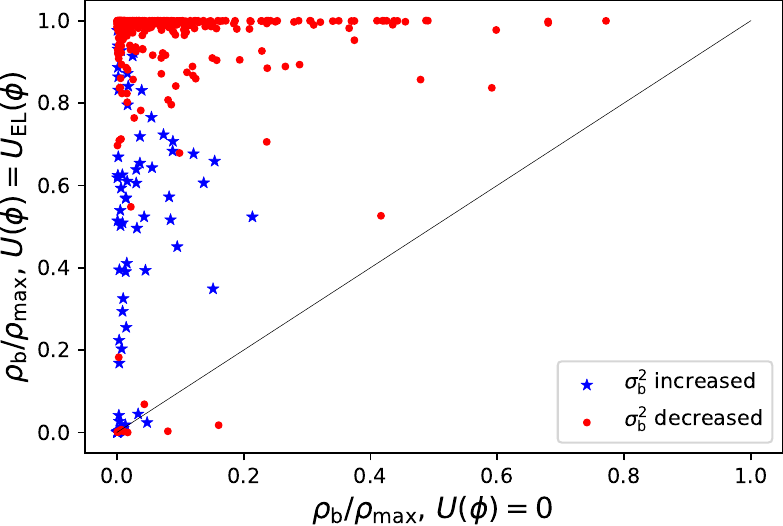}
   \label{fig:BianchiIX_EL_1.25_rho_compare_shear_marked}
\end{subfigure}
\begin{subfigure}
   \centering
   \includegraphics[width=0.49\linewidth]
   {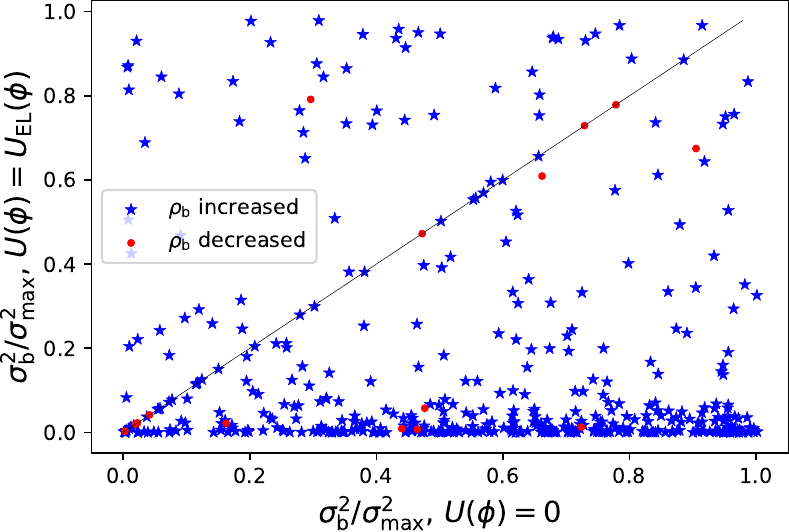}

   \label{fig:BianchiIX_EL_1.25_shear_compare_rho_marked}
\end{subfigure}

\begin{subfigure}
   \centering
   \includegraphics[width=0.49\linewidth]
   {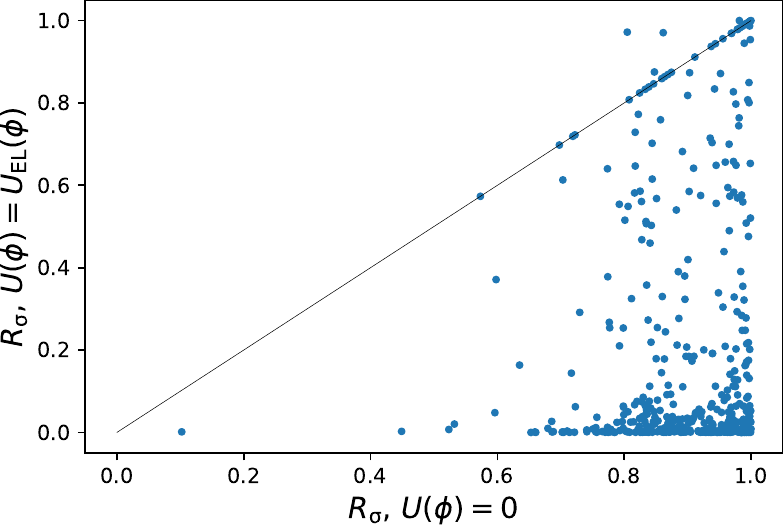}

\end{subfigure}

\caption{Bianchi-IX effective spacetime with an ekpyrotic-like field ($U_0 = 1.25$): energy density $\rho_\mathrm{b}/\rho_\mathrm{max}$, shear $\sigma_\mathrm{b}^2/\sigma_\mathrm{max}^2$, and the relative strength of anisotropies $R_\sigma$ at the last bounce compared for the ekpyrotic-like potential (y-axis) and the massless scalar case (x-axis) in 500 simulations with matching initial conditions. The panels follow the same color coding used in all previous comparison plots. Similar to the Bianchi-I case, strengthing the ekpyrotic potential does not yield drastic changes in isotropization. }

\label{fig:BianchiIX_EL_1.25}
\end{figure}

We now consider Bianchi-IX spacetimes with an ekpyrotic-like scalar field having $U_0=0.25.$ Fig. \ref{fig:BianchiIX_EL_0.25} shows the comparison of energy density, shear scalar, and the ratio $R_\sigma$ at the last bounce relative to these same values for the matching massless scalar field simulations. We find that, similar to the ekpyrotic potential simulations above, the ekpyrotic-like potential also causes isotropization at the bounce in Bianchi-IX spacetimes. This is especially visible in the final panel of Fig. \ref{fig:BianchiIX_EL_0.25} where we see that $R_\sigma$ decreases in $91.17\%$ of all simulations. Similar to the Bianchi-I case in LQC, we note that while this percentage is lower than under the ekpyrotic potential, the average value of $R_\sigma$ under the ekpyrotic-like potential ($0.26$ here) is actually smaller than under the ekpyrotic potential. This can be understood by comparing Fig. \ref{fig:BianchiIX_Cycle_0.25} with Fig. \ref{fig:BianchiIX_EL_0.25} and noticing that a larger portion of the simulations are clustered at the bottom of the plot in the ekpyrotic-like case.

As $R_\sigma$ captures the relative difference between the energy density and anisotropic shear, we also consider the effects of the ekpyrotic-like potential on these parameters in the first and second panels of Fig. \ref{fig:BianchiIX_EL_0.25}. In particular, the energy density is increased compared to the massless scalar field case in almost all simulations. However, this is again not necessarily accompanied by a decrease in the anisotropic shear. As in the case of the ekpyrotic potential in Bianchi-IX spacetimes in LQC, the first panel does not show the separation of simulations where the shear scalar increased (blue) and decreased (red) that was present in the Bianchi-I spacetimes in LQC. However, there are differences in the general behavior of these two cases, with a decrease in anisotropic shear most often accompanying simulations with a larger increase in energy density. 
The second panel considers the anisotropic shear, and the key point is that many of the simulations are clustered at the bottom of the figure. This is accompanied by $76.85\%$ of all simulations having a decrease in the anisotropic shear under the ekpyrotic-like potential compared with the massless scalar case. Hence, it is unsurprising that the average value of $\sigma_\mathrm{b}^2/\sigma_\mathrm{max}^2$ is $\sigma_\mathrm{av}^2 = 2.34$, a value smaller than any other case considered so far and consistent with the high degree of isotropization found under the ekpyrotic-like potential.

\subsubsection{Effective Bianchi-IX spacetime with ekpyrotic-like potential having $U_0=1.25$}

Finally, we  study the effect of increasing the ekpyrotic-like potential in Bianchi-IX spacetimes. This is studied in Fig. \ref{fig:BianchiIX_EL_1.25}, which compares the energy density, anisotropic shear, and ratio $R_\sigma$ under the ekpyrotic-like potential with $U_0=1.25$ versus under a massless scalar field. As in the Bianchi-I case, it is again surprising how similar Fig. \ref{fig:BianchiIX_EL_1.25} is compared to  Fig. \ref{fig:BianchiIX_EL_0.25}. This strongly contrasts with the ekpyrotic potential, where there are clear (if surprising) qualitative differences in the results due to the strengthening of the potential. In particular, we still note a strong isotropization effect of the ekpyrotic-like potential, which is indicated by both the energy density and the anisotropic shear panels. More specifically, we find a slight increase in isotropization under the strengthened potential, with $93.62\%$ of all simulations showing $R_\sigma$ smaller with the potential than without, and the average value of $R_\sigma$ dropping to $0.20$, the smallest value of any case considered in this paper.

\begin{table}
\centering

\begin{tabular}{|l|l|l|l|l|l|}
\hline
\multicolumn{1}{|r|}{{\ul \textit{\textbf{}}}}
&  &  &  & \textbf{\% of simulations} & \textbf{\% of simulations}    \\ 
& \textbf{$\rho_\mathrm{av}$} & \textbf{$\sigma_\mathrm{av}^2$} & \textbf{$R_{\sigma_\mathrm{avg}}$}  & \textbf{ where $\sigma^2$ decreased} & \textbf{where $R_\sigma$ decreased}    \\ \hline
\textbf{Bianchi-I: Massless Scalar Field}                 &            0.062 &               5.13 &              0.78 & N/A                                     & N/A                                 \\ \hline

\textbf{Bianchi-I: ekpyrotic $U_0=0.25$} &          0.226&               5.52&        0.49& 37.60\%& 100\%\\ \hline
\textbf{Bianchi-I: ekpyrotic $U_0=1.25$} &           0.361&               4.03&        0.27& 50.20\%& 100\%\\ \hline

\textbf{Bianchi-I: ekpyrotic-like $U_0=0.25$} &          0.320&               3.77&        0.29& 61.00\%& 97.40\%\\ \hline
\textbf{Bianchi-I: ekpyrotic-like $U_0=1.25$} &           0.332&               2.98&        0.25& 71.80\%& 96.00\%\\ \hline

\textbf{Bianchi-IX: massless scalar field}                 &            0.037&               6.52&              0.88& N/A                                     & N/A                                 \\ \hline

\textbf{Bianchi-IX: ekpyrotic $U_0=0.25$} &           0.303&               4.30&        0.34& 71.99\%& 93.97\%\\ \hline

\textbf{Bianchi-IX: ekpyrotic $U_0=1.25$} &           0.312&               6.53&        0.41& 56.26\%& 99.12\%\\ \hline

\textbf{Bianchi-IX: ekpyrotic-like $U_0=0.25$} &           0.316&               2.34&        0.26& 76.85\%& 91.17\%\\ \hline

\textbf{Bianchi-IX: ekpyrotic-like $U_0=1.25$} &           0.340&               1.88&        0.20& 81.06\%& 93.62\%\\ \hline
\end{tabular}
\caption{\textcolor{black}{The table summarizes results from various numerical simulations for Bianchi-I and Bianchi-IX spacetime in LQC. 500 simulations were performed for each case.}
Comparison of changes in $\rho$, $\sigma^2$ for matter fields with different potentials (the numbers represent the magnitude of $\rho$ and $\sigma^2$ at the bounce as a fraction of their respective maximum possible value in LQC). $R_\sigma$ is a measure of the relative strength of anisotropic shear with respect to energy density. 
\textcolor{black}{ The initial values of $c_1$ and $c_2$ are chosen in range $c_1, c_2 \in [-0.25,0]$, whereas the values of triads are chosen in range $p_1,p_2,p_3 \in [10000,30000]$. The value of $c_3$ is determined using the Hamiltonian constraint. The value of initial energy density is chosen in the range $\rho \in [0,10^{-4}]$. For $U_o = 0.25$, the range of $\phi$ lies in the range $\phi \in [0,0.0007] $ for the ekpyrotic potential \eqref{Ekpyrotic_potential}, whereas  
$\phi \in [0,0.4] $ for the ekpyrotic-like potential \eqref{potential_EL}. Also 
$\dot \phi \in [0,1.5 \times 10^{-2}]$ for the ekpyrotic potential and $\dot \phi \in [0,1]$ for the ekpyrotic-like potential. These values change slightly for $U_o = 1.25$.}
The values of $\rho$, $\sigma^2$ and $R_\sigma$ averaged over all 500 simulations for a given spacetime and potential configuration are denoted as $\rho_\mathrm{av}, \sigma^2_\mathrm{av}$ and $R_{\sigma_\mathrm{av}}$. Note that ekpyrotic and ekpyrotic-like potential with the same value of $U_o$ are not directly comparable as their corresponding potential strengths are quite different for the same values of $U_0$.}
\label{table:1}
\end{table}

$\phi \in [0,0.4] $ for the ekpyrotic-like potential \eqref{potential_EL}. 

\section{Summary}
Bouncing cosmologies where quantum gravitational effects lead to singularity resolution have been explored as potential scenarios to explain the conditions in the very early universe and are often discussed as potential alternatives to inflation. 
Earlier phenomenological studies in LQC strongly indicate that in the presence of ordinary matter, the Bianchi-I and Bianchi-IX effective spacetimes remain highly anisotropic after the bounce. Moreover, for such matter the bounce is more likely to increase anisotropies as anisotropies tend to dominate over matter in the bounce regime. This is primarily due to the fact that anisotropies grow at a much faster rate than ordinary matter when a contracting universe approaches the bounce. Thus, additional input is needed if isotropization is to be achieved.
In our analysis, we considered whether an ekpyrotic scalar field having an ultra-stiff equation of state, may not only help curb the growth of anisotropies but also enhance the isotropization of the resulting universe. 
Specifically, using the anisotropic bouncing cosmologies of LQC, we studied the problem of isotropization at the bounce with the help of two different ekpyrotic potentials, first in the effective Bianchi-I spacetime as the simplest anisotropic scenario, and then in effective Bianchi-IX spacetimes as the most generic anisotropic scenario including spatial curvature.  The intuitive reasons for expecting the ekpyrotic field to produce isotropization are easy to understand. While anisotropies in the bounce regime have an effective equation of state equivalent to that of stiff matter, the ekpyrotic field behaves as an isotropic fluid having an equation of state larger than unity. Thus, the ekpyrotic field is expected to grow faster than anisotropies as the universe contracts towards a bounce and produces isotropization \cite{Steinhardt:2002ih}. \textcolor{black}{Analysis of the singularity resolution coupled to ekpyrotic scenario in LQC has been carried out earlier, but the question whether ekpyrotic field is effective in causing isotropization of the bounce was not addressed.} To explore this, we consider the anisotropic Bianchi-I and Bianchi-IX models in LQC in the effective spacetime description as our setting to put the isotropizing ability of the ekpyrotic field to the test. Specifically, we study whether the ekpyrotic field can dominate the bounce regime and lead to a reduction in the strength of the anisotropic shear relative to the energy density of the universe (which is taken to be isotropic in this manuscript).  \textcolor{black}{It is important to note that the initial conditions are not specifically geared towards achieving an extended phase of ekpyrosis. The randomized initial conditions are chosen to be agnostic towards the occurrence and duration of a phase of ekpyrosis, which is left to be determined entirely by the underlying cosmological dynamics and the interplay between energy density and anisotropic shear. The objective of this work is to study how likely the ekpyrotic field is to overwhelm the anisotropic shear, starting from random initial conditions given in the classical macroscopic phase of the universe which is contracting. Thus, the initial conditions are not deliberately made favorable towards the ekpyrotic field.}

Through extensive numerical simulations, we studied the effect of ekpyrotic and ekpyrotic-like potentials and found that they have a strong isotropization effect on the universe. We found that compared to the massless scalar field as the matter content of the universe, the inclusion of these potentials generically decreased $R_\sigma$, our measure of anisotropy which indicates the relative strength of the anisotropic shear compared to the isotropic energy density, in over $90\%$ of cases; regardless of which effective spacetime or potential was considered. \textcolor{black}{It should be noted that in our numerical simulations, the initial conditions of the scalar field and its velocity were small which are not ideal for ekpyrosis to occur. This is the consequence of theoretical upper bounds on energy density and anisotropic shear in LQC. Even with such not favorable initial conditions we find that sufficient isotropization is achieved.  We find that even short-lived phases of ekpyrosis coupled with multiple bounces are sufficient to cause the isotropization for the initial conditions considered in our work. This indicates that for a different quantization scheme allowing larger bounds on energy density and anisotropic shear, the ekpyrotic phase is expected to be stronger.} Importantly, our simulations indicate that this reduction in $R_\sigma$ is primarily achieved by an enhancement of the isotropic energy density at the bounce. Thus, while the energy density at the bounce is increased compared to massless scalar field in almost all cases, the effect on the anisotropic shear is not as decisive.

A comparison of the effects of the two ekpyrotic potentials in Bianchi-I and Bianchi-IX effective spacetimes is given in Table \ref{table:1}.  The simulations often resulted in a succession of bounces before the emergence of a semi-classical universe. Since we are interested in isotropization at the bounce as a mechanism for isotropization of the large-scale universe, we attempted to pick time intervals of simulations so that a semi-classical expanding universe is finally obtained, and the results reported in Table \ref{table:1} are for the last bounce. However, the extremely complex behavior of the dynamical equations of the effective spacetime, particularly for Bianchi-IX universe, led to difficulties in choosing such a time scale for the simulations. We also studied the effect of increasing the strength of the potentials, indicated by the potential parameter $U_0$. As shown in the table, increasing the strength of the potential often led to higher isotropization, though not always. We also note that isotropization is achieved in more cases in the Bianchi-I spacetime compared to the Bianchi-IX spacetime. This is expected as the presence of spatial curvature in the Bianchi-IX spacetime complicates the interplay between anisotropic shear and energy density.

Several aspects of our results merit further inspection. An immediate question stems from the lack of direct correlation between the reduction obtained in the ratio $R_\sigma$ and the effect on the anisotropic shear. This requires an understanding of the complex relation between the energy density and the anisotropic shear in the bounce regime. While this has recently been partially understood in the case of Bianchi-I spacetimes where the shear scalar and energy density were found to satisfy a simple relationship at the bounce, no such insights are yet available for the Bianchi-IX models. Already, the changes in the clustering of points between Figures \ref{fig:BianchiIX_Cycle_0.25} and \ref{fig:BianchiIX_Cycle_1.25} hint towards some interesting relationship in the case of Bianchi-IX spacetimes. This question will be explored in future work. Another question, immediate from the setup of the problem, is how our results depend on the spacetimes and quantization methods considered.

It is also important to understand the role of quantization ambiguities in this study. One of the questions will be to consider 'A' versus 'F' quantization in Bianchi-IX spacetimes which utilizes different quantizations of the spatial curvature \cite{Singh:2011gp}. Finally, we notice various oddities, such as the increased strength of the ekpyrotic-like potential yielding a surprisingly small impact on the results when compared to the ekpyrotic potential. Moreover, the probability of isotropization (indicated by the percent of simulations where $R_\sigma$ decreased) and the degree of isotropization (indicated by the decrease in the average value of $R_\sigma$ compared to the massless scalar field case) does not seem to be simply related and requires further investigation. Further study of the effects of varying the potential may help illuminate answers to these questions.

Finally, we note that our results on the efficacy of ekpyrosis in producing isotropization point towards an isotropization mechanism which is expected to hold in general for all bouncing models. The primary mechanism through which ekpyrosis achieves isotropization is by significantly enhancing the contribution of matter energy density in the approach to the bounce, thus reducing the relative strength of anisotropies. In our analysis we find that ekpyrosis affects the anisotropies only indirectly through the equations of motion, and a change in the absolute magnitude of anisotropic shear is at best a secondary contributory factor in the isotropization effect produced by ekpyrosis. Thus, the specific way in which the equations of motion are modified in the bounce regime, which determines how the ekpyrotic field is related to the anisotropic shear, plays a secondary role in isotropization. Hence, even though the specific bounce mechanism used in this manuscript is the one provided by LQC, we expect the results on the efficacy of ekpyrosis in achieving isotropization to be robust and to hold for generic bouncing scenarios. We show that ekpyrosis has a strong isotropization effect on the universe in bouncing models. Moreover, our results show that the strength of ekpyrotic potentials is expected to be positively correlated to the isotropization achieved. Further, by comparing the results obtained in the effective dynamics of the Bianchi-IX model with those from the effective dynamics of the Bianchi-I model, we show that the presence of spatial curvature along with anisotropies does not significantly hamper the efficacy of ekpyrosis. In other words, ekpyrosis continues to have a strong isotropization effect on the universe even in presence of spatial curvature.

\section*{Acknowledgments}
\textcolor{black}{We thank the anonymous referee for suggestions which improved the presentation in the manuscript.}
R.B and P.S. are supported by NSF grants PHY-2110207 and PHY-2409543. A.M.M thanks the REU Site in Physics
and Astronomy (NSF Grant No. 1852356) at Louisiana State University where a major part of this work was performed.  A.M.M is supported by the Perimeter Institute for Theoretical Physics. Research at Perimeter Institute is supported in part by the Government of Canada through the Department of Innovation, Science and Economic Development and by the Province of Ontario through the Ministry of Colleges and Universities. \textcolor{black}{S.S. is supported by Guru Jambheshwar University of Science \& Technology, India with Seed Money Grant vide Endst. No. DR\&D/2025/103-10.}

\bibliographystyle{h-physrev5}

\end{document}